\RequirePackage{fix-cm}
\documentclass[twocolumn,epjc3]{svjour3} 

\smartqed  
\RequirePackage{graphicx}
\RequirePackage{units}
\RequirePackage{booktabs}
\RequirePackage{amsmath}
\RequirePackage{makecell}
\RequirePackage{subfigure}
\RequirePackage[colorlinks,citecolor=blue,urlcolor=blue,linkcolor=blue]{hyperref}
\journalname{Eur. Phys. J. C}
\def\makeheadbox{\relax} 

\begin{document}

\thankstext{c1}{Corresponding author, mukund@mpp.mpg.de}
\thankstext{c2}{Corresponding author, rituparna.maji@oeaw.ac.at}
\thankstext{c3}{Corresponding author, martin.stahlberg@mpp.mpg.de}
\thankstext{c4}{Corresponding author, alex.stendahl@helsinki.fi}

\institute{
Max-Planck-Institut f\"ur Physik, 85748 Garching - Germany \label{mpp}
\and
Institut f\"ur Hochenergiephysik der \"Osterreichischen Akademie der Wissenschaften, 1050 Wien - Austria \label{hephy}
\and
Atominstitut, Technische Universit\"at Wien, 1020 Wien - Austria \label{ati}
\and
CNR-SPIN c/o Dipartimento di Scienze Fisiche e Chimiche, Universit\`a degli Studi dell'Aquila, 67100 L'Aquila - Italy \label{unilaq_cnr}
\and
Consiglio Nazionale delle Ricerche, Istituto di Nanotecnologia, 00185 Roma - Italy  \label{cnr}
\and
Department of Mathematics and Physics, North Carolina Central University, 27707 Durham NC - USA \label{nccu_phys}
\and
Department of Nuclear Engineering, North Carolina State University, 27695 Raleigh NC - USA \label{nccu_nuc}
\and
Department of Physics, Duke University, 27708 Durham NC - USA \label{duke}
\and
Department of Physics, ETH Zurich, CH-8093 Zurich, Switzerland \label{eth_physics}
\and
Dipartimento di Scienze Fisiche e Chimiche, Universit\`a degli Studi dell'Aquila, 67100 L'Aquila - Italy \label{unilaq}
\and
ETH Zurich - PSI Quantum Computing Hub, Paul Scherrer Institute, CH-5232 Villigen, Switzerland \label{eth_quantum}
\and
Gran Sasso Science Institute, 67100 L'Aquila - Italy \label{gssi}
\and
Helsinki Institute of Physics, 00014 University of Helsinki - Finland \label{hip}
\and
INFN - Laboratori Nazionali del Gran Sasso, 67010 Assergi - Italy \label{infn_lngs}
\and
INFN - Sezione di Roma, 00185 Roma - Italy \label{infn_roma}
\and
SNOLAB, P3Y 1N2 Lively - Canada \label{snolab}
\and
State Key Laboratory of Functional Crystals and Devices, Shanghai Institute of Ceramics, Chinese Academy of Sciences, 201899 Shanghai - P.R.C \label{siccas}
\and
Triangle Universities Nuclear Laboratory, 27708 Durham NC - USA \label{tunl}
}

\author{
G.~Angloher\thanksref{mpp}
\and
M.~R.~Bharadwaj\thanksref{c1,mpp}
\and
A.~Böhmer\thanksref{hephy,ati}
\and
M.~Cababie\thanksref{hephy,ati}
\and
I.~Colantoni\thanksref{cnr,infn_roma}
\and
I.~Dafinei\thanksref{gssi,infn_roma}
\and
N.~Di~Marco\thanksref{gssi,infn_lngs}
\and
C.~Dittmar\thanksref{mpp}
\and
L.~Einfalt\thanksref{hephy,ati}
\and
F.~Ferrella\thanksref{gssi}
\and
F.~Ferroni\thanksref{gssi,infn_roma}
\and
S.~Fichtinger\thanksref{hephy}
\and
A.~Filipponi\thanksref{unilaq,infn_lngs}
\and
T.~Frank\thanksref{mpp}
\and
M.~Friedl\thanksref{hephy}
\and
M.~Gapp\thanksref{mpp}
\and
L.~Gai\thanksref{siccas}
\and
Z.~Ge\thanksref{siccas}
\and
M.~Heikinheimo\thanksref{hip}
\and
M.~N.~Hughes\thanksref{mpp}
\and
K.~Huitu\thanksref{hip}
\and
M.~Kellermann\thanksref{mpp,hephy,ati}
\and
R.~Maji\thanksref{c2,hephy,ati}
\and
M.~Mancuso\thanksref{mpp}
\and
L.~Pagnanini\thanksref{gssi,infn_lngs}
\and
F.~Petricca\thanksref{mpp}
\and
S.~Pirro\thanksref{infn_lngs}
\and
F.~Pr\"obst\thanksref{mpp}
\and
G.~Profeta\thanksref{infn_lngs,unilaq}
\and
A.~Puiu\thanksref{infn_lngs}
\and
F.~Reindl\thanksref{hephy,ati}
\and
K.~Sch\"affner\thanksref{mpp}
\and
J.~Schieck\thanksref{hephy,ati}
\and
P.~Schreiner\thanksref{hephy,ati}
\and
C.~Schwertner\thanksref{hephy,ati}
\and
K.~Shera\thanksref{mpp}
\and
M.~Stahlberg\thanksref{c3,mpp}
\and
A.~Stendahl\thanksref{c4,hip}
\and
M.~Stukel\thanksref{snolab,infn_lngs}
\and
C.~Tresca\thanksref{unilaq_cnr,infn_lngs}
\and
F.~Wagner\thanksref{hephy,eth_physics,eth_quantum}
\and
S.~Yue\thanksref{siccas}
\and
V.~Zema\thanksref{mpp,hephy}
\and
Y.~Zhu\thanksref{siccas}
\and
P.S.~Barbeau\thanksref{duke,tunl}
\and
S.C.~Hedges\thanksref{duke,tunl}
\and
C. Awe\thanksref{duke,tunl}
\and
J. Runge\thanksref{duke,tunl}
\and
T. Johnson\thanksref{duke,tunl}
\and
D.M. Markoff\thanksref{nccu_phys}
\and
P. An\thanksref{duke,tunl}
\and
C. G. Prior\thanksref{duke,tunl}
\and
A. Bracho\thanksref{duke,tunl}
\and
S. Alawabdeh\thanksref{nccu_nuc}
}

\title{Quenching factors for Na recoils as a function of Tl dopant concentrations in NaI(Tl) crystals}

\date{\today}

\maketitle

\begin{abstract}
 Thallium-doped sodium iodide (NaI(Tl)) scintillation detectors play an important role in the field of direct dark matter (DM) searches. The DAMA/LIBRA experiment stands out for its reported observation of an annually modulating DM-like signal, which is in direct contrast with other results.
 To accurately calibrate the energies of nuclear recoil signals with electron recoils, precise measurements of the quenching factor of the NaI(Tl) crystals are essential, as the two processes have different scintillation light yield.
 In this article, we present results of a systematic study carried out by the COSINUS collaboration and Duke University to measure the quenching factor of sodium (Na) recoils as a function of nuclear recoil energy and for differing Thallium (Tl) dopant concentrations in the bulk crystal. 
 Five ultrapure NaI(Tl) crystals, manufactured by the Shanghai Institute for Ceramics, were irradiated with a quasi-monoenergetic neutron beam at the Triangle Universities Nuclear Laboratory, North Carolina, USA.
 The quenching factor for low nuclear recoil energies of \unit[5-26]{keV$_{\textrm{nr}}$} was extracted for all 5 crystals. A Tl-dependence could be deduced with a proportional response calibration schema using a $^{241}$Am source. However, this effect was not observed when using a low-energy calibration line from $^{133}$Ba.
 
\end{abstract}

\section{Introduction}
\label{introduction}

The utilization of Thallium-doped sodium iodide (NaI(Tl)) crystals as particle detectors has been a cornerstone in nuclear, medical and particle physics since their discovery in 1949 \cite{nai_discovery}.
These detectors are notable for their high intrinsic scintillation light yield and the ease of growing large-size crystals, making them appealing for a wide variety of experimental applications. Among them is the attempt at direct detection of dark matter (DM) \cite{bernabei2022recent,Fushimi_2016,sabre,PhysRevD.103.102005,PhysRevD.106.052005,PhysRevD.95.032006}.
Despite their historical significance, the precise measurement of NaI(Tl) detectors’ response to nuclear scattering events remains an ongoing area of research.

The DAMA/LIBRA experiment’s observation of an annually-periodic modulation signal in the low-energy region of NaI(Tl) detectors, potentially indicating DM interactions, has heightened the importance of resolving these measurement challenges \cite{bernabei2022recent}. While other direct DM search experiments using different target materials have reported null results \cite{billard2022direct}, ongoing experiments like ANAIS-112 \cite{PhysRevD.103.102005,Coarasa2024ANAIS112TY}, COSINE-100 \cite{PhysRevD.106.052005,Carlin_2025_v2} and COSINUS \cite{Angloher_2016}, along with planned efforts from SABRE \cite{sabre} and PICOLON \cite{Fushimi_2016} aim to provide a cross-check of the DAMA/LIBRA results using the same target material. The combined datasets of ANAIS-112 and COSINE-100 are in $4.7\sigma$ and $3.5\sigma$ tension with DAMA/LIBRA in the \unit[1-6]{keV} and \unit[2-6]{keV} regions respectively \cite{Carlin_2025}, assuming spin-independent elastic WIMP scattering off the target material.

Calibration in NaI(Tl)-based DM experiments is typically performed using gamma ($\gamma$)-rays interacting with electrons in the crystal, establishing the electron-equivalent energy scale (E$_{\textrm{ee}}$). For heavier particles, such as neutrons, less energy is converted into light for the same initial recoil energy. The quenching factor (QF) quantifies this difference, and is needed to calculate the nuclear recoil energy E$_{\textrm{nr}}$ from the measured light signal. Therefore, the interpretation of signals and the WIMP parameter space coverage in experiments relying on scintillation-only detection heavily depend on the estimation of the QF.
Previous measurements have revealed discrepancies, particularly at lower energies \cite{1,2,3,4,5,6,7,8,9,10,11}, which can significantly influence the interpretation of results from DM experiments.
These discrepancies may arise from factors such as crystal doping, manufacturing processes, and non-linearities in light output.
Recent results \cite{10,11} show how the non-linear effects of scintillation light can lead to systematic shifts in the reported QF at lower energies. In addition, data processing plays a role, e.g. via the low-energy trigger efficiency and noise control near threshold regions, as noted in \cite{6,11}.
A different approach is employed by the COSINUS experiment, which uses undoped NaI crystals as cryogenic scintillating calorimeters \cite{christmas_run}.
This method provides a direct measurement of the recoil energy via a phonon signal, which is nearly independent of the interacting particle type. As the scintillation light is also measured using a second channel, the QF can be deduced in-situ \cite{summer_run}.

This article presents a systematic examination of the sodium (Na) QF in five radio-pure NaI(Tl) crystals, with a focus on its energy dependent behavior in the low-energy recoil regime from \unit[5-26]{keV$_{\textrm{nr}}$}.
Sec. \ref{setup} describes the experimental setup and the data-acquisition system (DAQ) used for the QF measurement.
Sec. \ref{analysis} covers the data processing, energy calibration and event selection, while Sec. \ref{simulation} describes the simulation framework used.
Sec. \ref{qf} brings together the simulated data and the experimental data to estimate the QF for Na recoils. Both energy dependence and the influence of Tl(ppm) dopant content are investigated.

\section{Experimental setup}
\label{setup}
\subsection{Overview}
The measurements were conducted at the Advanced Neutron Calibration Facility at the Triangle Universities Nuclear Laboratory (TUNL) in Durham, North Carolina.
The quasi-monoenergetic neutrons that are required for QF measurements are produced using a tandem Van de Graff accelerator.
Protons were created in a pulsed beam with a pulse period of \unit[400]{ns} and a timing resolution of \unit[2]{ns} using a Direct Extraction Negative Ion Source (DENIS) \cite{denis}.
A \unit[1434]{nm} Lithium Fluoride (LiF) foil evaporated onto a tantalum substrate placed at the target location was irradiated by the beam. The nominal incident proton energy was \unit[3]{MeV}.
Resultant quasi-monoenergetic neutrons with an energy of \unit[$\sim$1260]{keV} (with a small spread due to proton energy loss in LiF as seen in Fig. \ref{Simulation/neutron_source}) are produced via the $^7\textrm{Li}(p,n)^7\textrm{Be}$ reaction \cite{LISKIEN197557,FRIEDMAN2013117}. 
The beam energy and LiF thickness were configured to generate recoil energies of interest at relatively small scattering angles, while simultaneously ensuring a sufficient event rate for collecting adequate statistics for the measurement without pile-up concerns.
Calculations were performed to determine the overall interaction rate in the NaI(Tl) crystal at low recoil energies, utilizing the $^7\textrm{Li}(p,n)^7\textrm{Be}$ cross-sections \cite{LISKIEN197557,FRIEDMAN2013117} and the differential neutron elastic scattering cross-sections for $^{23}$Na and $^{127}$I at the relevant angles.
\begin{figure}[t]
    \centering
    \includegraphics[width=0.44\textwidth]{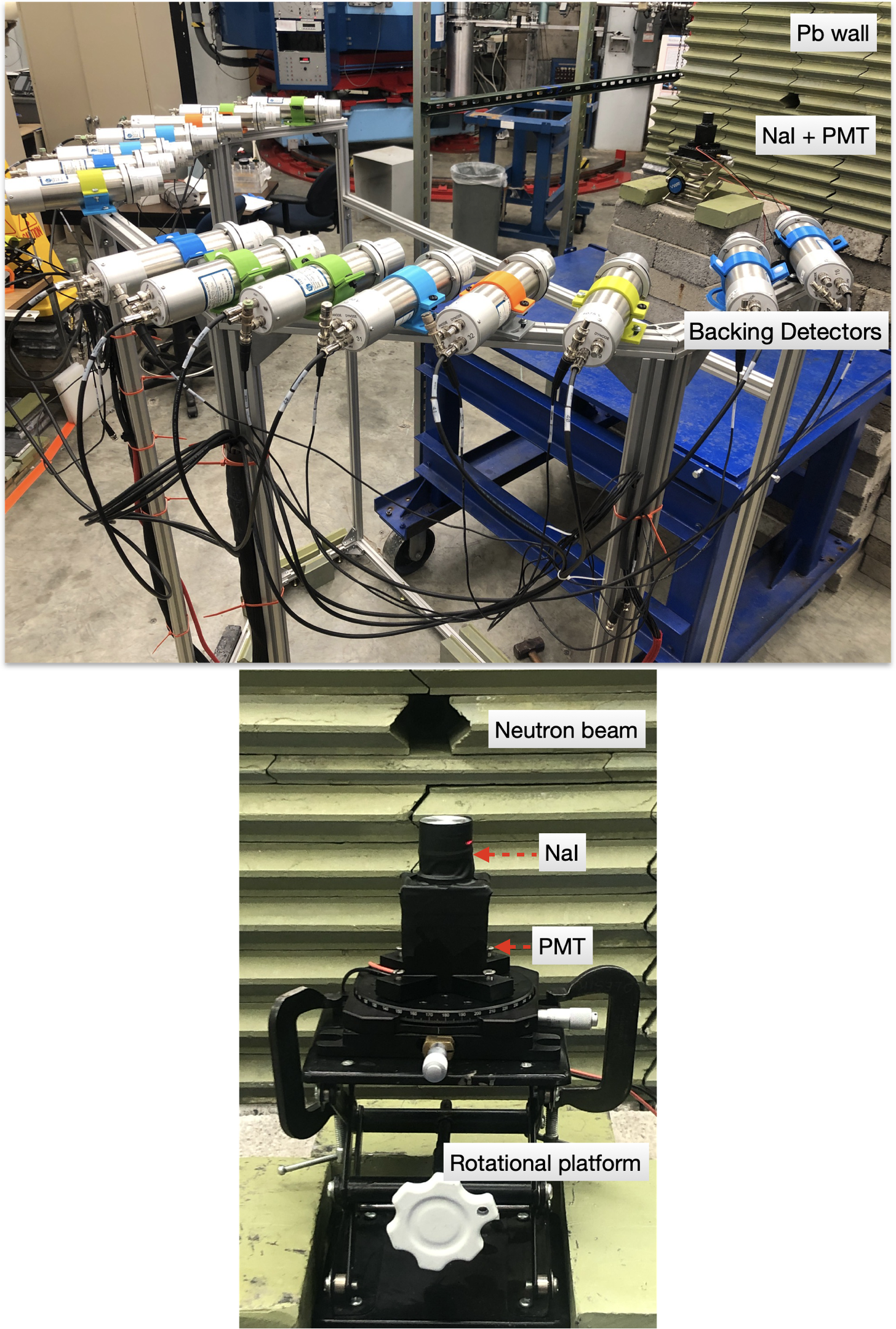}
    \caption{Top: The experimental setup at TUNL; Bottom: A close-up of the NaI(Tl) crystal coupled to the PMT.}
    \label{setup_tunl}
\end{figure}
A Beam Pulse Monitor (BPM) signal was used as a timing reference, marking when the pulsed proton beam interacted with the LiF target to produce the neutrons.

Bi-layer shielding consisting of high density Polyethylene (HDPE) and borated HDPE was placed around the enclosure of the LiF target to block any off-axis neutrons, and a collimated slit was used to direct the outgoing beam towards the NaI(Tl) crystal.
The resultant beam had an angular spread of \unit[2$\pm$0.3]{$^\circ$}. 
An additional layer of lead (Pb) (\unit[10]{cm} thick) covered the front surface to reduce the fraction of secondary $\gamma$-rays produced by the neutron capture of hydrogen in the HDPE. The setup is pictured in Fig. \ref{setup_tunl}

\subsection{Detector configuration}

The NaI(Tl) crystals were manufactured at the Shanghai Institute for Ceramics of the Chinese Academy of Sciences using a modified Bridgman technique described in Ref. \cite{siccas}.
``Astro-Grade" powder from Merck Co. (previously Sigma Aldrich) \cite{merck} was used as the starting material.
Inductively coupled plasma mass spectrometry (ICP-MS) measurements performed at LNGS \cite{LNGS} showed contamination at a level of \unit[6-22]{ppb} for $^{40}$K and $<$ \unit[1]{ppb} for $^{232}$Th and $^{238}$U, respectively. 
Overall, five samples were prepared, with Tl dopant levels of 0.1, 0.3, 0.5, 0.7 and \unit[0.9]{\%} respectively in the initial powder. 
The Tl dopant levels in the cut crystals were found to vary from \unit[284]{ppm} to \unit[1390]{ppm} as listed in Table \ref{runtime} with an uncertainty of \unit[20]{\%} from the ICP-MS method.

\begin{table}[h]
    \centering
    \footnotesize
    \renewcommand{\arraystretch}{1.2}  
    \setlength{\tabcolsep}{2pt}  
    \begin{tabular}{c c c c c} 
        \toprule
        \textbf{Crystal} & \textbf{\makecell{Tl conc. \\ (powder)}} & \textbf{\makecell{Tl conc. \\ (cut crystal)}} & \textbf{Run time} \\ [0.4ex] 
        \midrule
        1  & 0.7\% & 1270 $\pm$ 254 ppm & 32h 16min \\ 
        2  & 0.3\% & 425 $\pm$ 85 ppm & 28h 34min \\
        3  & 0.9\% & 1390 $\pm$ 278 ppm & 35h 57min \\ 
        4  & 0.1\% & 284 $\pm$ 56 ppm & 33h 43min \\ 
        5  & 0.5\% & 800 $\pm$ 160 ppm & 35h 27min \\ 
        \bottomrule
    \end{tabular}
    \caption{Tl-dopant concentration, for both powder and cut crystal, and run time for each NaI(Tl) crystal.}
    \label{runtime}
\end{table}

The synthesized crystals are right-circular cylinders with a uniform height and diameter of \unit[2.54]{cm} each, respectively.
Given the hygroscopic nature of NaI(Tl), each crystal was encased in an aluminum enclosure with a thickness of \unit[1.25]{mm}. 

The housing material, in combination with the small size of the crystal, was chosen to minimize the possibility of multiple scatters.
Assembly was carried out inside a nitrogen-flushed glovebox to ensure the crystals did not degrade.
A photo-multiplier tube (PMT) manufactured by Hamamatsu Photonics (weighing \unit[74]{g}; Model number: H11934-200) was optically coupled to the bottom face of the crystal (using EJ-550 silicone grease) to measure the light output.
This setup was covered with a layer of black adhesive tape before being mounted on a custom 3D printed motorized unit that could be tuned to rotate slowly about its axis to minimize the impact of any channeling effects during data-taking, as seen in Fig. \ref{setup_tunl}.
Each crystal to be tested was placed at a distance of \unit[1.016]{m} from the LiF target in line with the beam axis.

Liquid scintillation detectors, which we denote as backing detectors (BD), were installed to tag the neutrons scattered off the Na or I nuclei. The BD contain a cylindrical EJ-309 liquid scintillation cell with a diameter and length of \unit[2]{inches}, manufactured by Eljen Technology \cite{eljen}, coupled to a Hamamatsu R7724 PMT.
Previous measurements \cite{9,10} effectively demonstrated the excellent pulse-shape discrimination capability of these detectors, allowing for accurate tagging of neutron events.
In total, 15 BDs were deployed for this run, with their corresponding scattering angles and distance from the NaI(Tl) crystal provided in Table \ref{bdangles}.

The aforementioned BD angles were chosen to constrain our region of interest below \unit[30]{keV$_\textrm{{nr}}$} for Na recoils in the current study.
Each of the BDs was encapsulated inside a \unit[1.5]{mm} thick Al housing and equipped with a Pb shielding cap in front and around their enclosure during operation to reduce the background gamma trigger rate.
An additional BD was employed as a time-of-flight (TOF) detector to measure and monitor the spread of the neutron beam energy.

For the alignment of the entire setup, a laser range finder was used.
A set of reference points were used for lining up the NaI(Tl) crystal to the center of the beam, which was checked after each rotation and crystal swap.
The range finder was used to align all the BDs on the same plane as that of the NaI(Tl) crystal and the beam line.

\begin{table}[h]
    \centering
    \footnotesize 
    \renewcommand{\arraystretch}{1.1}  
    \setlength{\tabcolsep}{5pt}
    \begin{tabular}{c c c} 
        \toprule
        \textbf{Detector no.} & \makecell{\textbf{Scattering} \\ \textbf{angle}} & \makecell{\textbf{Distance} \\ \textbf{(cm)}} \\ [0.5ex] 
        \midrule
        BD0   & $40^\circ$ & 101.5 \\ 
        BD1   & $35^\circ$ & 115.7 \\
        BD2   & $27.5^\circ$ & 131.6 \\
        BD3   & $22.5^\circ$ & 136.1 \\
        BD4   & $18^\circ$ & 142.0 \\
        BD5   & $14.5^\circ$ & 145.9 \\
        BD6   & $11^\circ$ & 152.5 \\
        BD7   & $7^\circ$ & 154.3 \\
        BD8   & $9^\circ$ & 155.3 \\
        BD9   & $12^\circ$ & 151.1 \\
        BD10  & $15.5^\circ$ & 145.9 \\
        BD11  & $22^\circ$ & 139.0 \\
        BD12  & $27^\circ$ & 131.3 \\
        BD13  & $32.5^\circ$ & 123.5 \\
        BD14  & $37.5^\circ$ & 107.8 \\
        TOF   & $0^\circ$ & 195.7 \\ [1ex] 
        \bottomrule
    \end{tabular}
    \caption{Scattering angles and distances for each backing detector. The measuring uncertainty on the angles is negligible considering the surface area of the BD.}
    \label{bdangles}
\end{table}

\subsection{Electronics and DAQ}
For data acquisition, a pair of SIS3316 14-bit digitizers with a sampling rate of 250 MHz manufactured by Struck Innovative Systeme was used to capture the data from all 15 BDs, the TOF detector, the BPM, and the NaI(Tl) PMT whenever one of the BD triggered.
The clocks of the two digitizers are synchronized, and the trigger is handled by an external NIM module.
As the expected energies to be measured in the BDs are far above their respective detection threshold, the analysis thresholds for BDs were not optimized individually. Instead, the gains of the BDs were simply adjusted to be similar and avoid dark counts/noise triggers.
Communication with the digitizers is carried out via the VME backplane using a Struck 3150 VME interface.
The acquisition control is managed by the NGMDAQ software package, developed at Oak Ridge National Laboratory \cite{ngmdaq}.

\subsection{Measurement summary}
The measurement was carried out by the TUNL group over two weeks in September 2021, with cumulative run times for each crystal presented in Table \ref{runtime}. The NaI(Tl) crystals were calibrated with $^{133}$Ba and $^{241}$Am sources at the beginning and end of each individual run respectively. For the calibration datasets, the trigger scheme was changed to record events in the NaI(Tl) PMT.

\section{Data Analysis}
\label{analysis}

\subsection{Data processing}
\begin{figure}[h]
    \centering
    \includegraphics[width=0.45\textwidth]{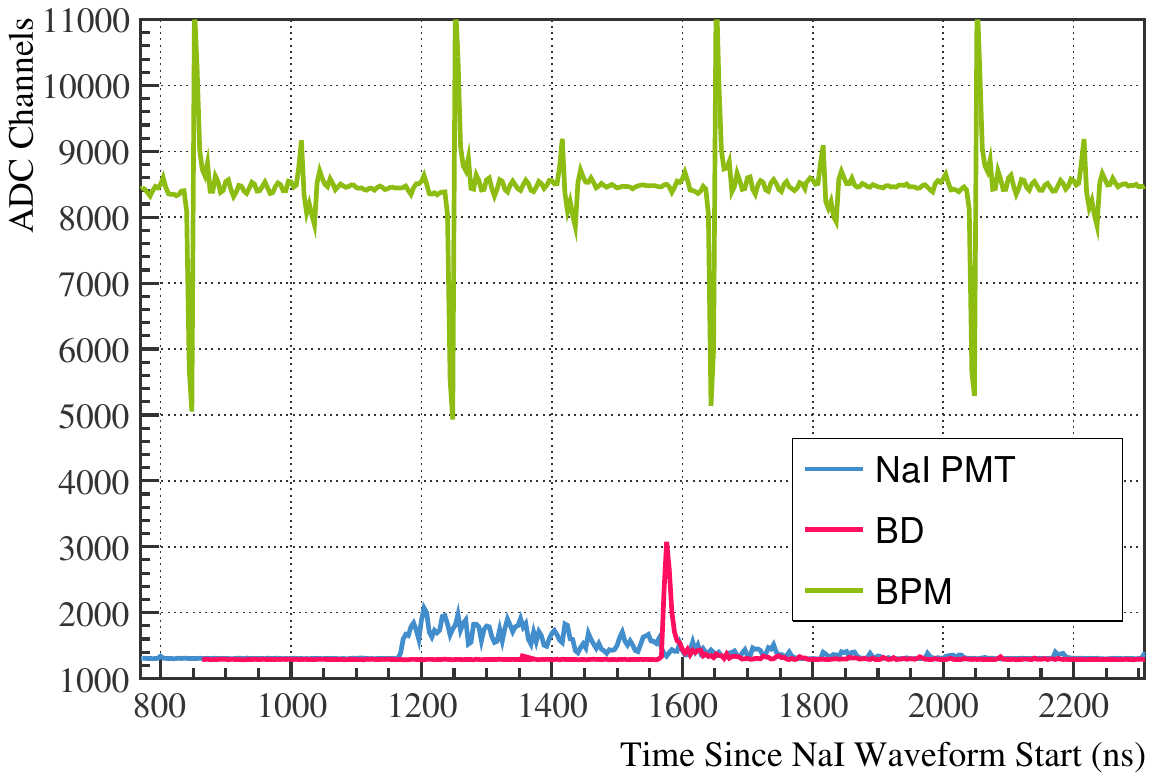}
    \caption{Measured pulse shapes from the PMT attached to the NaI(Tl) crystal and BD 1 for a nuclear recoil event of $\sim\unit[60]{keV_{\textrm{ee}}}$. The BPM provides a shaped signal as timing reference for identifying beam-induced neutron events.}
    \label{triggering}
\end{figure}

A coincidence mechanism was employed between the NaI(Tl) detector and each of the BDs to select events of interest in the NaI(Tl) detector. For each coincidence event, the contributing BD was determined by comparing the pulse onset timing information with the coincidence trigger time, which was $\sim$\unit[350]{ns}, depending on the position of the BD.
The typical time difference between the prior BPM pulse and the NaI(Tl) pulse onset was found to be $\sim$\unit[240]{ns}. It should be noted here that the BPM signal has an arbitrary offset to the NaI and BD signals. A typical beam-induced nuclear recoil event is illustrated in Fig. \ref{triggering}. 

Using an estimate of the time window in which NaI pulses occur within the record, a fixed window integration pulse reconstruction scheme was employed with an integration window of \unit[2.8]{\textmu s}. In line with measurements reported by other studies \cite{9}, it was observed that the low-energy NaI(Tl) events could not be well reconstructed with this method due to integrating the PMT noise, which becomes dominant in this regime. Therefore, the charge estimation from \cite{9} was implemented as an energy estimator instead, where only samples above a certain threshold are integrated. The threshold was set to \unit[11]{ADC} units, slightly below the single photonelectron (SPE) charge at \unit[12.2]{ADC} units. This reconstruction method was used for the whole energy range, as no nonlinearity with respect to the default integration was observed at higher energies.


\subsection{Calibration}
\label{calibration}

\begin{figure}[h]
    \centering
    \includegraphics[width=.47\textwidth]{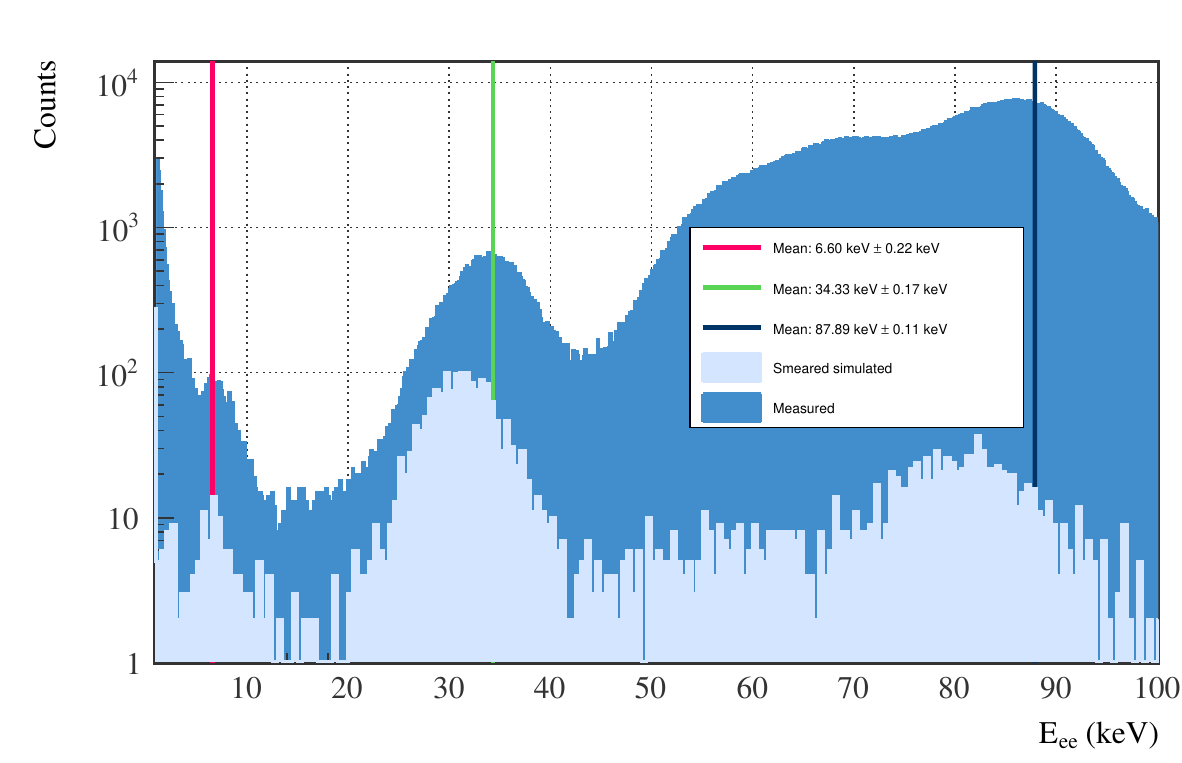}
    \caption{Observed and simulated spectrum in \unit[]{keV} from the $^{133}$Ba calibration of Crystal 1. The measured spectrum was calibrated using the \unit[6.6]{keV} proportional response. The simulated spectrum is smeared by Gaussian convolution, where the width is given by the resolution function $\sigma(E) = a\sqrt{E}$. The measured peaks are fitted with a Gaussian on top of a linear background. The \unit[6.6]{keV} line is close to the noise peak, then in order of increasing energy the peaks are identified as the unresolved sum of \unit[30.62]{keV} and \unit[30.97]{keV}, and finally the \unit[81.0]{keV} line. The vertical lines denote the fitted Gaussian mean converted to energy using the \unit[6.6]{keV} proportional response. Deviations of the measured spectra from simulation are expected since the simulation does not account for the nonlinearity of the scintillation response in NaI(Tl) and the measured spectrum is calibrated using a proportional response.}
    \label{ba133_cal}
\end{figure}

Two different calibration methods were chosen to study the impact of different calibration schemes on the final QF estimation.
A linear, single-point calibration using the \unit[59.54]{keV} $^{241}$Am $\gamma$-peak serves as a standard cross-check with previously reported measurements.
A second single-point calibration using a $^{133}$Ba source was applied to evaluate the effect of using a low-energy peak to calibrate the energy scale. Fig. \ref{ba133_cal} shows the observed and the simulated Monte-Carlo spectrum for the $^{133}$Ba calibration of Crystal 1.
From the simulated spectrum, the feature close to the noise peak was attributed to the escape of $^{127}$I K$_{\alpha}$ X-rays following photoelectric absorption of the \unit[35.1]{keV} K$_{\beta}$ X-rays from $^{133}$Ba in the NaI(Tl) crystal. 
A mean energy of \unit[6.6]{keV} was obtained from the simulation, and used for calibration. 
The scintillation light response of NaI(Tl) generally exhibits an energy-dependent non-linearity \cite{det_cal_1,det_cal_2,refId0}. Here, this effect is not accounted for, as it cannot be modeled precisely. From Fig. 5 in \cite{refId0}, the relative difference in LY between \unit[50]{keV} and \unit[6]{keV} is expected to be $\sim$10\%, which explains the discrepancy between the measured and simulated peaks at higher energies in Fig. \ref{ba133_cal}. Similarly, the linear calibration with \unit[59.54]{keV} from $^{241}$Am would systematically overestimate low measured energies. In our region of interest ($\sim$\unit[0.5]{keV} $< E_{\textrm{ee}} <$ \unit[5]{keV}), the relative LY is expected to decrease by another $\sim$10\%.

\subsection{Identification of nuclear recoils}
\label{nucleartags}
A set of Pulse Shape Discrimination (PSD) and timing cuts is applied to the events triggered by the BDs to select only scattered neutron events, while rejecting accidental triggers due to ambient/scattered coincident gammas. The PSD cut is based on the charge comparison method, using the ratio of the deposited charge sum of the tail (Q$_\textrm{tail}$) to the total charge (Q$_\textrm{tot.}$) in the BD signal.
The integration time window for the tail was \unit[340]{ns}, while the total integration time window was \unit[400]{ns}.

\begin{figure}[h]
    \centering
    \begin{subfigure}
        \centering
        \includegraphics[width=.4\textwidth]{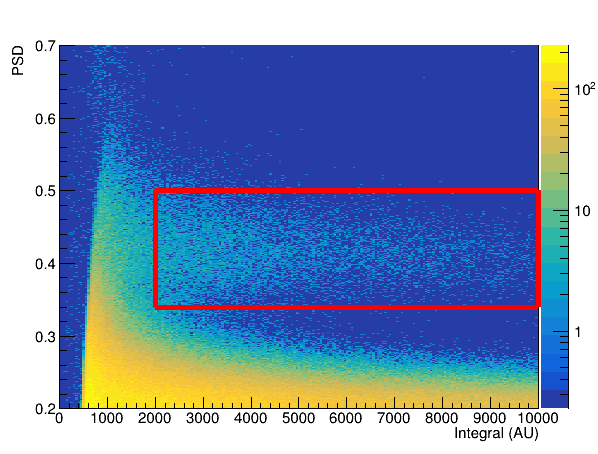}
    \end{subfigure}
    \hfill
    \begin{subfigure}
        \centering
        \includegraphics[width=.4\textwidth]{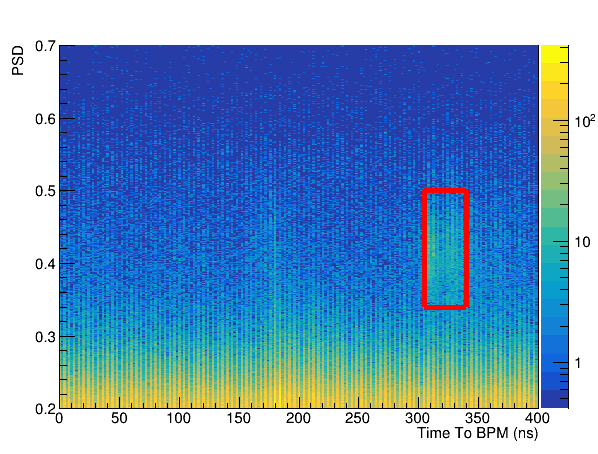}
    \end{subfigure}
    \caption{PSD cut (top) and Time to BPM cut (bottom) applied to triggered events of BD 0 for Crystal 1.}
    \label{cuts}
\end{figure}

\begin{figure}[h]
    \centering
        \centering
        \includegraphics[width=.49\textwidth]{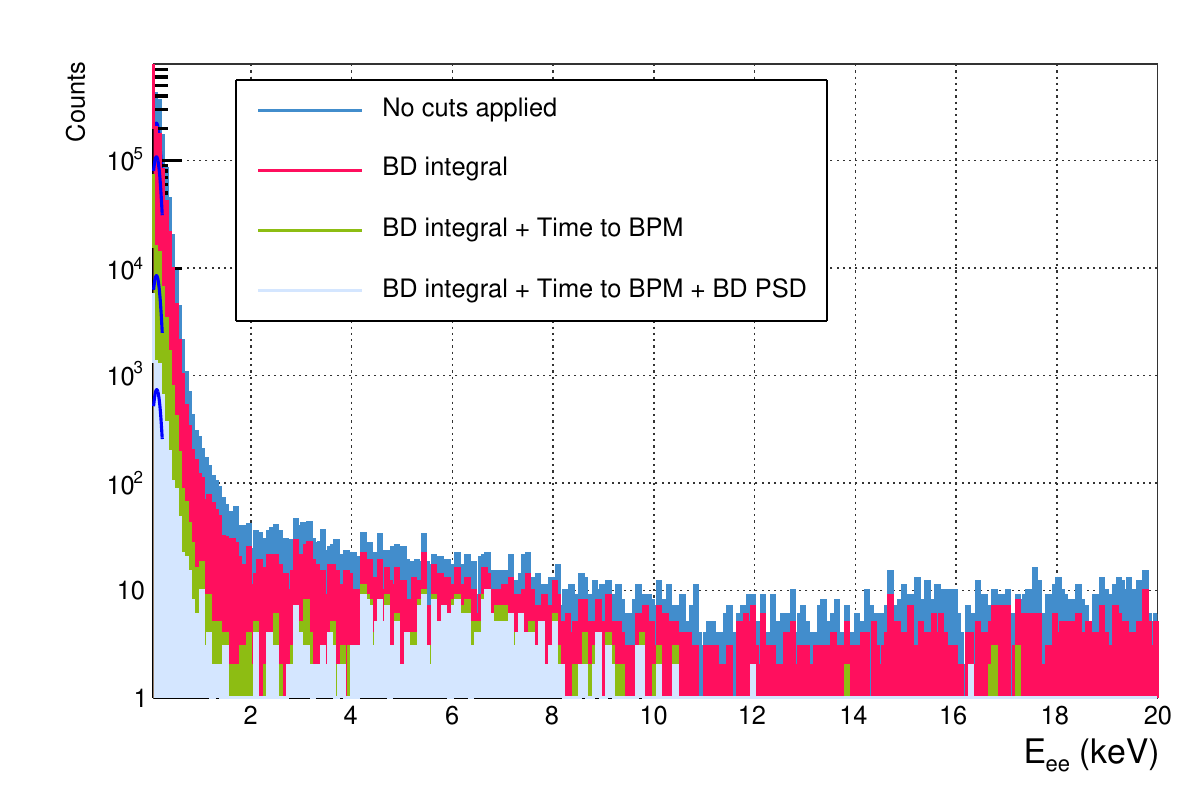}
    \caption{Measured PMT spectra for Crystal 1, triggered by BD0. The impact of each cut parameter is shown, where the values are illustrated in Fig. \ref{cuts}.}
    \label{uncut}
\end{figure}

The TOF cut (with respect to the BPM signal) removes accidental triggers resulting from neutrons that may have scattered off different parts of the experimental setup and not off the crystal.
This is particularly helpful as the TOF of the scattered neutrons from the Na or I nuclei to the BD is almost constant for an incident monoenergetic neutron beam (with only a small spread arising from the initial neutron energy
distribution). A final cut on the deposited charge in the BDs was applied to reduce noise triggers. The cuts applied to Crystal 1 and BD0 are shown in Fig. \ref{cuts} while Fig. \ref{uncut} shows the spectrum before and after applying the highlighted cuts.


\section{Simulation}
\label{simulation}

\subsection{Simulated geometry}
\label{Sec.simulated_geometry}
An extensive Monte Carlo simulation was performed with the Geant4~\cite{GEANT4:2002zbu,Allison2006Geant,ALLISON2016186} based software ImpCRESST \cite{CRESST:2019oqe}, initially developed for the CRESST DM search experiment~\cite{abdelhameed2019first}. An ImpCRESST version equipped with Geant4 v10.7.3 and ROOT v6-22-08~\cite{brun1997root} was used for this work. 
The primary goal of the simulation was to determine the distribution of nuclear recoil energies deposited in the NaI(Tl) crystal from scattered neutrons reaching each of the BDs. Additionally, the simulation was used to generate the energy spectrum resulting from interactions with the $^{133}$Ba and $^{241}$Am sources, which are employed for energy calibration as outlined in Sec. \ref{calibration}. Moreover, the simulation aimed to optimize the angular positioning parameters for the BDs to ensure each BD is exposed to a sufficient neutron flux within the energy range of interest. The results of the ImpCRESST simulation include detailed information on particle trajectories, energy deposition, particle types, interaction processes, and time of interaction.
\begin{figure}[t]
    \centering
    \includegraphics[height=5.7cm,width=8cm]{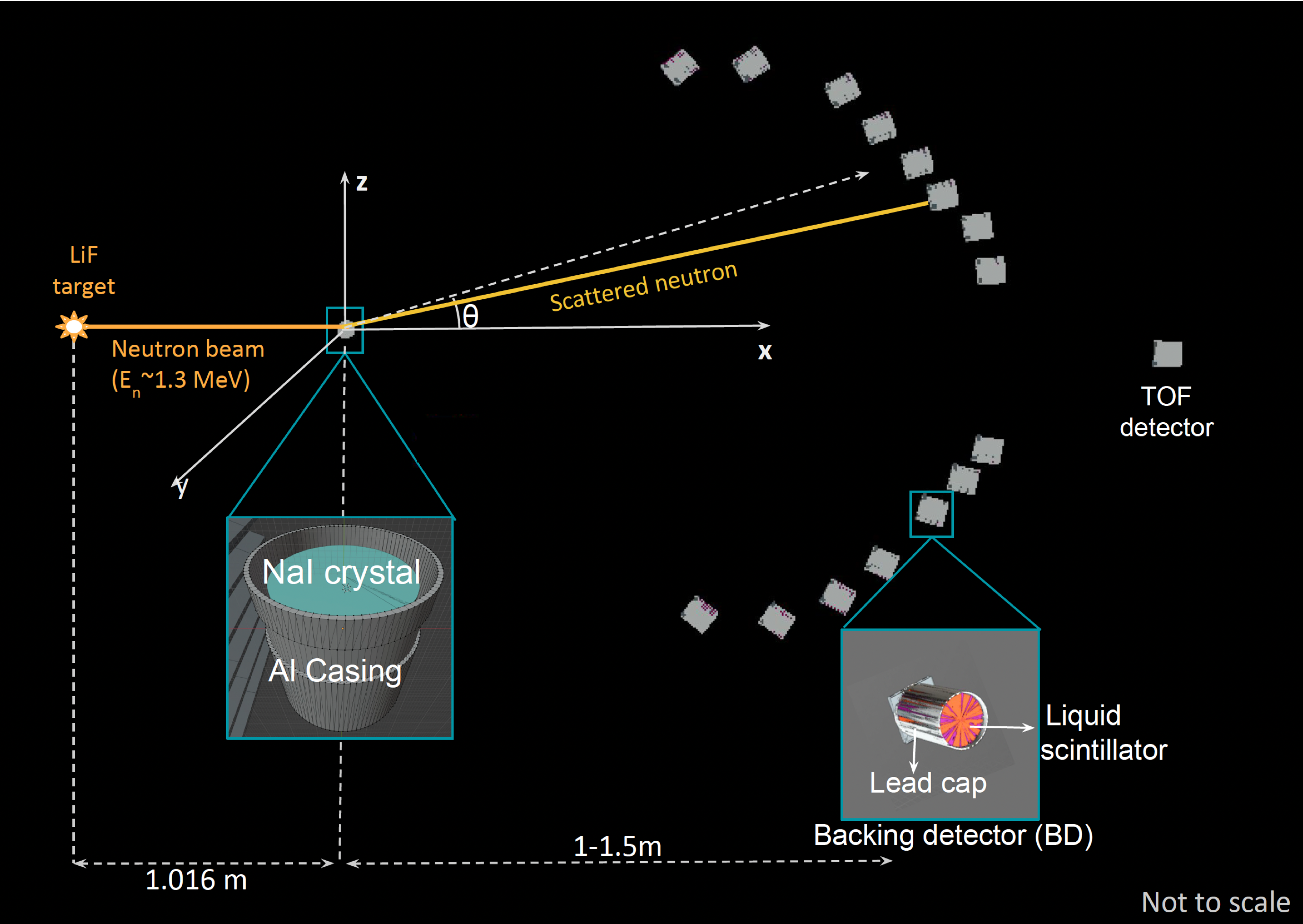}
    \caption {Schematic of the Geant4-simulated setup.}
    \label{Simulation/Schematic_setup}
\end{figure}

The modeled geometry outlined in Fig. \ref{Simulation/Schematic_setup} comprises a neutron beam produced at the LiF target, directed at the NaI(Tl) crystal housed in an Al casing, the array of 15 BDs, placed at the angles from Table \ref{bdangles}, and the separate TOF detector at 0$^\circ$ relative to the neutron beam. Each BD consists of a liquid scintillator (EJ-309) enclosed in a cylindrical Al housing and shielded by a Pb cap.

\subsection{Simulation results}
The amount of energy deposited in each experimental volume along with the respective time is recorded for each simulated neutron event. Two separate simulations were conducted, one under neutron irradiation and the other using three different gamma calibration sources. 

\subsubsection{Simulation of gamma calibration of the NaI(Tl) crystal}

Due to the absence of a precise description of the encapsulation for both the $^{133}$Ba and $^{241}$Am sources, they were simulated as point sources and positioned at the top of the NaI(Tl) crystal in the simulation. Fig. \ref{ba133_cal} shows the measured calibration and the corresponding simulated spectrum for the $^{133}$Ba source. The peak energies from the simulation were employed in the calibration process described in Sec. \ref{calibration}.

\subsubsection{Simulation of nuclear recoil energy distribution in NaI(Tl)}
\label{e_nr}
\begin{figure}[h]
    \centering
    \includegraphics[width=.49\textwidth]{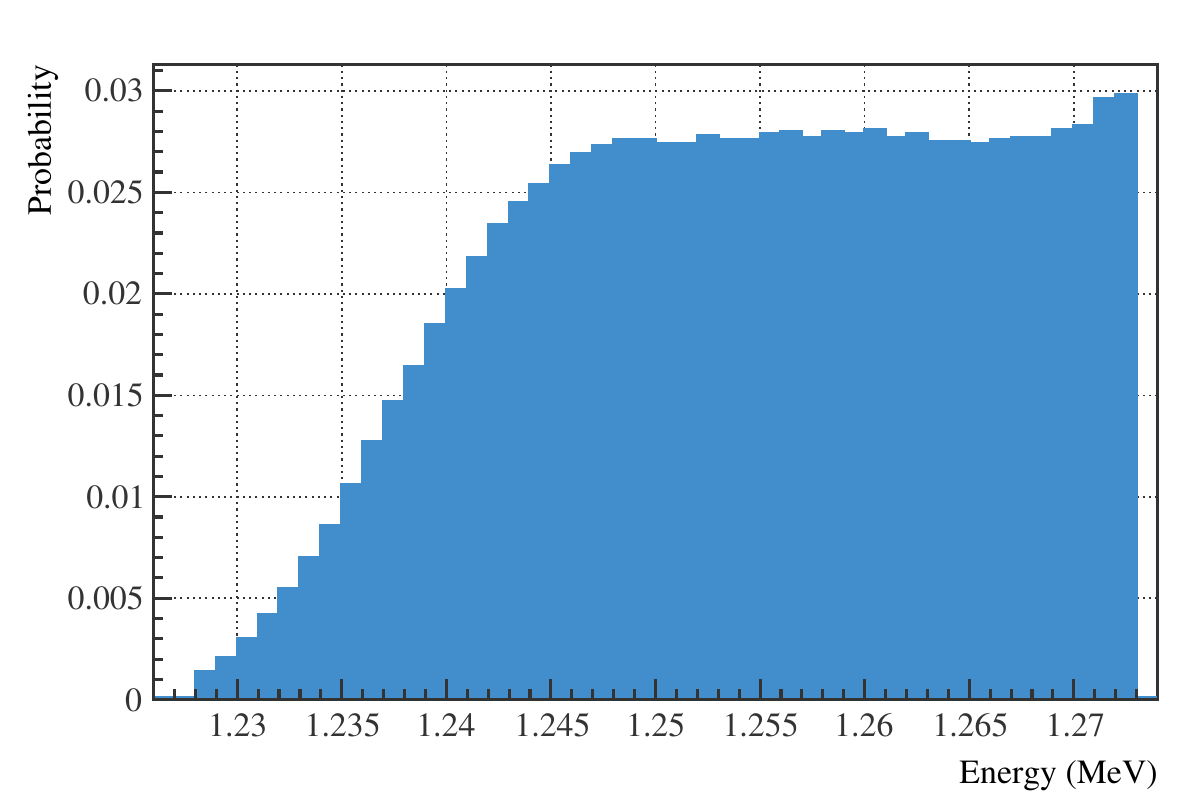}
    \caption {Normalized neutron energy spectrum used in the simulation, obtained from TRIM simulations tuned to the measurement from the TOF detector.}
    \label{Simulation/neutron_source}
\end{figure}

To determine the true nuclear recoil energy for each BD, the simulated geometry was irradiated by a quasi-monoenergetic neutron beam, treated as a point-like source, emitting neutrons confined to a narrow \unit[2.356]{$^\circ$} angular range. The input neutron spectrum was determined in the following way: using neutron production cross sections in the LiF target from \cite{LISKIEN197557,FRIEDMAN2013117} and the TRIM software package \cite{srim}, different spectra were simulated for different monoenergetic incident protons, and subsequently compared to the measured data from the TOF detector for neutron events. The best-fit incident proton energy is \unit[2969]{keV}. Subsequently, \unit[12]{B} neutrons were simulated using the input energy spectrum shown in Fig. \ref{Simulation/neutron_source}. 

For each neutron simulation, the energy deposition, interaction time, interaction type, and the involved particles in each detector are recorded.
If a scattered neutron is detected by one of the BDs after interacting with the NaI(Tl) detector, the event is tagged as a triggered event.
The nuclear recoil energy depends on the scattering angle of the neutrons, as described by Eq. \ref{Eqn_E_nr}:

\begin{align}\label{Eqn_E_nr}
    E_{nr} = \,&2E_{n} \frac{m_{n}^{2}}{(m_{n} + m_{N})^{2}}  \times  \nonumber\\
    &\left( \frac{m_{N}}{m_{n}} + \sin^{2}\theta - \cos\theta \sqrt{\frac{m^2_{N}}{m^2_{n}} - \sin^{2}\theta} \right)
\end{align}

where $E_n$ is the incident neutron energy, $\theta$ is the neutron scattering angle, $m_n$ is the neutron mass, and $m_N$ is the target nuclide mass.
For each neutron interaction in the NaI(Tl) crystal that triggers any BD, the interaction type (elastic or inelastic) and the specific nucleus involved (Na or I) are recorded. The simulation generates a distribution of the nuclear recoil energies in the NaI(Tl) crystal coinciding with any BD.
\begin{figure}[h]
    \centering
    \includegraphics[width=.48\textwidth]{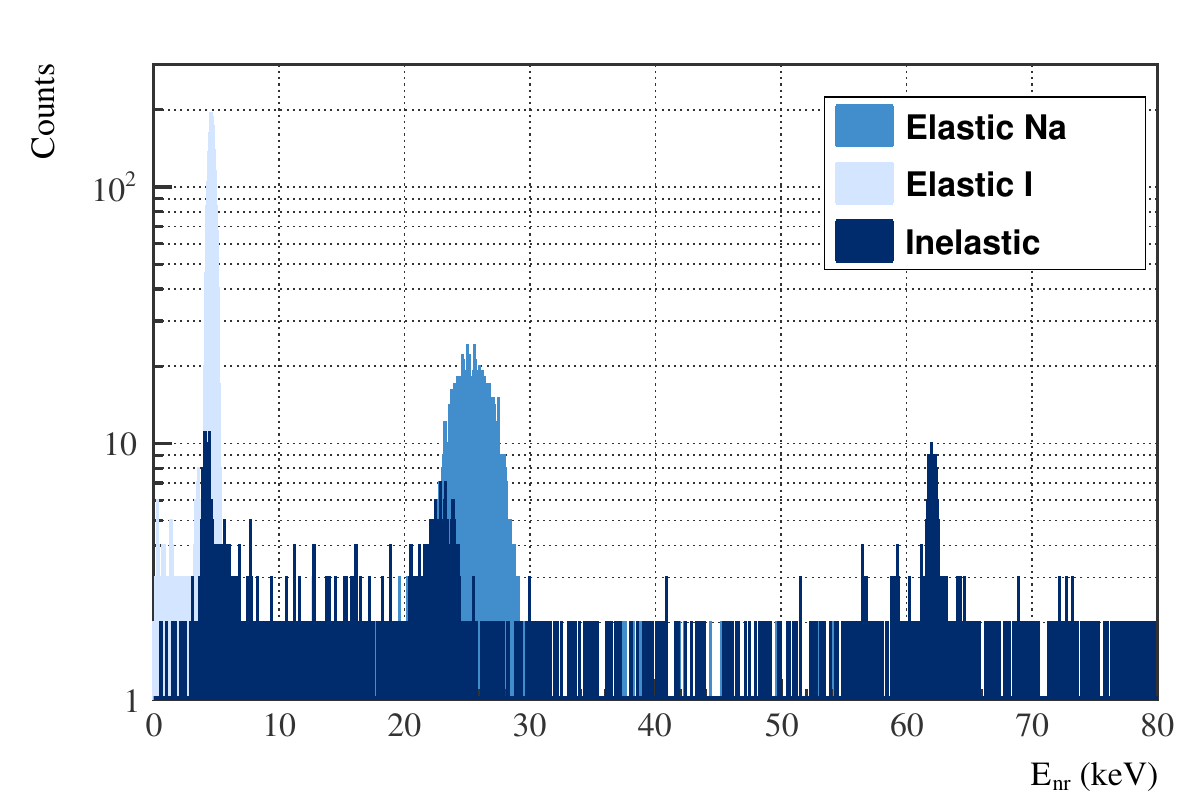}
    \caption {Simulated nuclear recoil energy deposition in the NaI(Tl) crystal, triggered by BD0. The elastic scattering off $^{23}$Na and $^{127}$I are shown in blue and light blue, while the inelastic scattering events are shown in dark blue.}
    \label{Simulation/Simulated_E_nr}
\end{figure}

The resulting nuclear recoil energy deposited in the NaI(Tl) crystal for BD0 is shown in Fig. \ref{Simulation/Simulated_E_nr}.
The simulation accounts for uncertainties in detector positions, neutron beam angular spread, and detector dimensions, as well as potential backgrounds from multiple neutron scattering within the setup. The multiple scattering events form a roughly constant background with small statistics and are not shown for simplicity.


\section{Quenching Factor estimation}\label{qf}

\label{bat_description}
The calibrated and cleaned energy spectra for nuclear recoils, expressed in the E$_{\textrm{ee}}$ energy scale obtained in Sec. \ref{calibration} and  Sec. \ref{nucleartags}, are used together with the simulated nuclear recoil spectra obtained in Sec. \ref{e_nr}, expressed in the E$_{\textrm{nr}}$ scale, to extract the quenching factor for Na recoils (QF$_{\textrm{Na}}$) for different energies for each crystal. 
For the highest recoil energy, corresponding to events from BD0, the quenched energy for I recoils is expected to be only \unit[$\sim$ 0.5]{keV$_{\textrm{ee}}$}, which is close to the background and noise pedestal in our measurement. Since we would obtain at most one data point for QF$_{\textrm{I}}$, we do not include I recoils in the fit function to reduce the number of free parameters.

\subsection{Signal and background modeling for QF extraction}

To account for the differing background components in each crystal and BD, data with no beam as well as beam data with an empty detector housing in the beamline was recorded. However, the observed background distributions in these datasets differ strongly from those observed in datasets with beam and NaI crystals. Therefore, we instead define a parametric background function that can vary freely between crystals and BDs.
The background count is modeled as \( BG(E) = ae^{bE} + cE^{d} \), with the free parameters $a,b,c,d$. 
The signal (Na recoils) is modeled by ``quenching" the simulated elastic recoil energy spectra (cf. Fig. \ref{Simulation/Simulated_E_nr}) with the QF, and smearing the resulting histograms with a resolution function. Each bin center from the simulated histograms is first multiplied with the QF, which is a free parameter per BD and crystal. This gives the quenched histogram bin centers $E_{q,j}$. For a given measured (i.e., quenched) energy value $E$, the value of the signal function $S(E)$ is then calculated by summing over the contributions \(S_j\) from each simulated histogram bin:

\begin{align}
    S(E) = A\sum_j^{N_\textrm{bins, sim}} S_j \bigg[ & \text{Erf}\left(E+\frac{w}{2},E_{q,j}, \sigma(E_{q,j})\right) \nonumber \\
    & - \text{Erf}\left(E-\frac{w}{2},E_{q,j}, \sigma(E_{q,j})\right) \bigg]
    \label{bin_count_eq}
\end{align}
\label{bin_content}
where \(S_j\) is the bin content of the simulated histogram bin \(j\), \(w\) is the simulated histogram bin width, \(\sigma(E_{q,j})\) the resolution evaluated at the quenched bin center, and \(\text{Erf}\) is the error function. A parameter controlling the overall number of signal counts (\(A\)) is also included. The recoil-peak energy resolution was parameterized as \(\sigma(E) = a\sqrt{E}\).
The final smeared signal can be asymmetric around a central value, but the individual bin smearings are Gaussian distributed, which is a reasonable assumption as long as the binning of the original simulated histogram is fine enough.
This signal is then added on top of the BG function to calculate the overall counts per bin. Example fits to the nuclear recoil spectrum measured for Crystal 1 with BD0 and BD2 are shown in Fig. \ref{fig:fit_bd0} and Fig. \ref{fig:fit_bd2} respectively. The fit range is \unit[0.133]{keV} to \unit[17.9]{keV}.\\
We fit the combined model using the Bayesian Analysis Toolkit (BAT) \cite{Caldwell_2009}, ensuring that all related uncertainties in the QF analysis chain are properly accounted for and propagated.
BAT is a Markov Chain Monte Carlo (MCMC) toolkit that explores the full available parameter space. The Markov chain is then used to create posterior distributions, from which the global mode and uncertainties on the QF are obtained.
All parameters have flat prior distributions, except for the resolution function. Here, the parameter $a$ is first fitted independently for each crystal and BD. Using the modes and associated uncertainties, we construct crystal-specific Gaussian priors for $a$ by computing an uncertainty-weighted average across BDs for each crystal; the combined propagated uncertainty defined the prior width. The analysis was then repeated with these priors to obtain the final results. 

\begin{figure}[h]
    \centering
    \includegraphics[width=.45\textwidth]{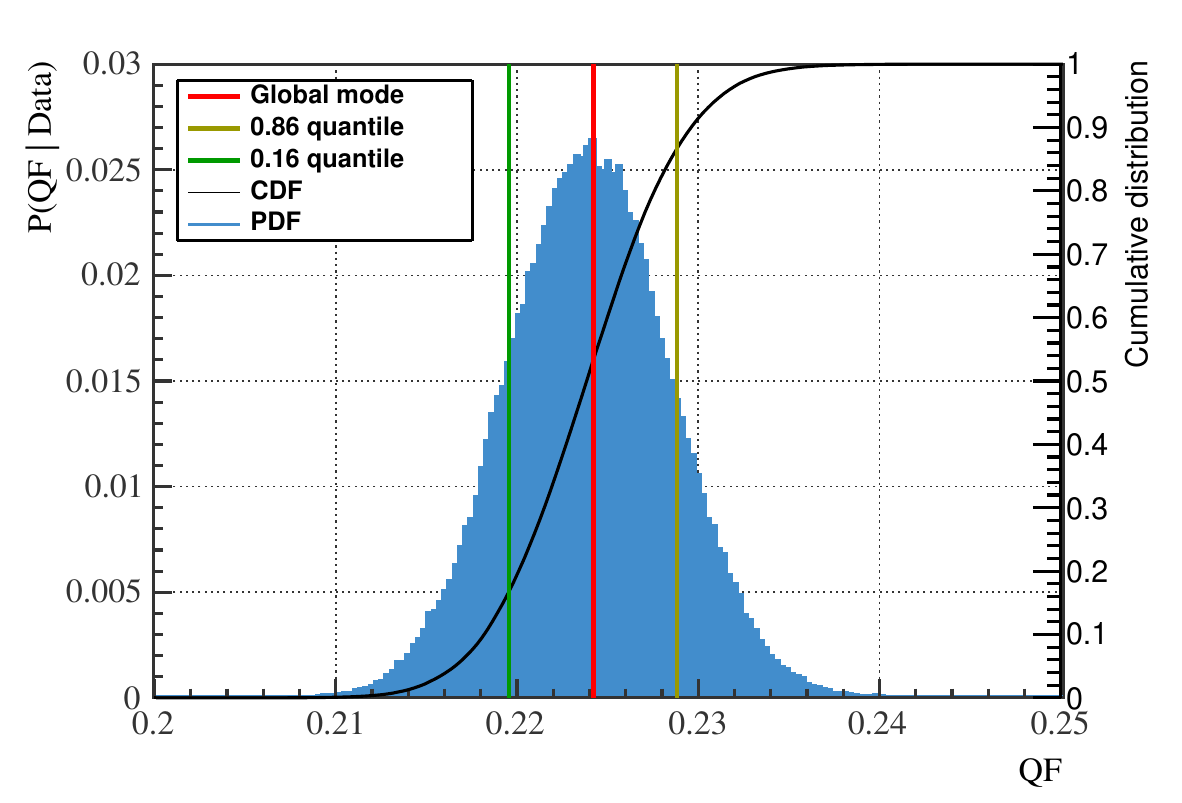}
    \caption{Posterior distribution for the QF for Crystal 1 BD 0, using the $^{133}$Ba proportional response, and the $\sigma = a\sqrt{E}$ resolution function. The green and yellow lines mark the 16\% and 86\% quantiles, the red is the global mode.}
    \label{fig:posterior}
\end{figure}

\begin{figure}[h]
    \centering
    \includegraphics[width=.49\textwidth]{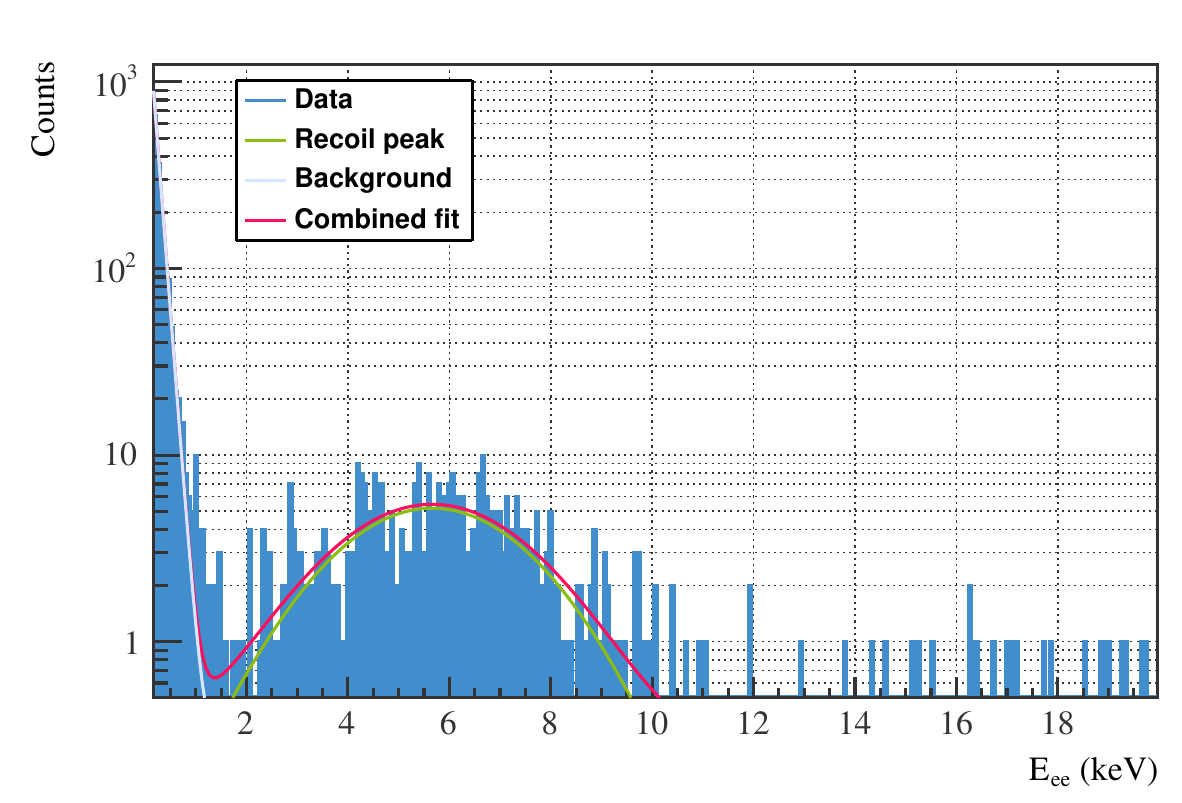}
    \caption{Fit of the recoil peak and background to the dataset for Crystal 1 and BD0.}
    \label{fig:fit_bd0}
\end{figure}

\begin{figure}[h]
    \centering
    \includegraphics[width=.49\textwidth]{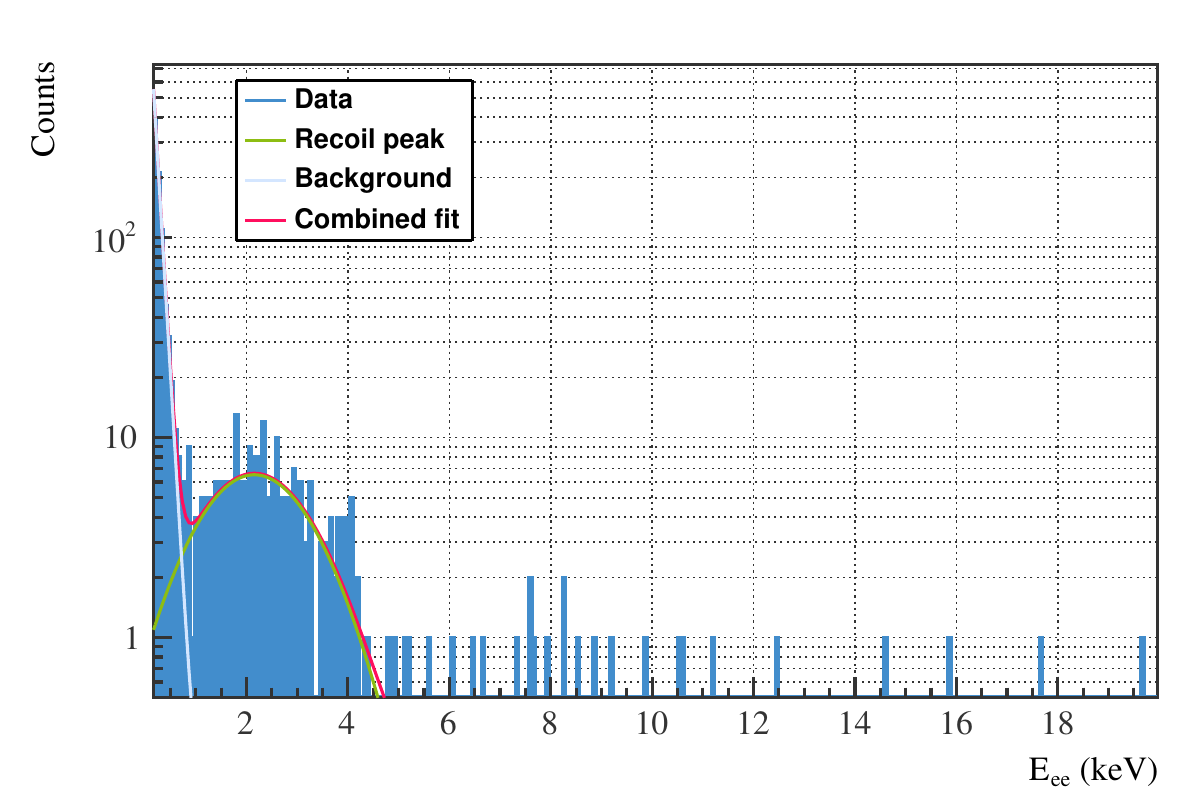}
    \caption{Fit of the recoil peak and background to the dataset for Crystal 1 and BD2.}
    \label{fig:fit_bd2}
\end{figure}

Convergence of the MCMC chains is evaluated by the $\hat{R}$-parameter \cite{brooks_gelman}, and the step size is optimized to ensure a chain efficiency between 15-35\%. The burn-in has an upper limit of 100,000 samples, but is stopped as soon as convergence is reached and all adjustable parameters meet their requirements. This condition was met in general with $\mathcal{O}(50,000)$ samples. We then run 5 chains for 200,000 samples to construct the posterior distribution.
The upper and lower uncertainties are derived from the 0.16 and 0.86 distribution quantiles, with the global mode as the central value.
Fig. \ref{fig:posterior} shows the resulting posterior distribution for the QF$_{\textrm{Na}}$ parameter, which appears Gaussian and centered around the global mode.


\begin{figure}[h]
    \centering
    \includegraphics[width=0.47\textwidth]{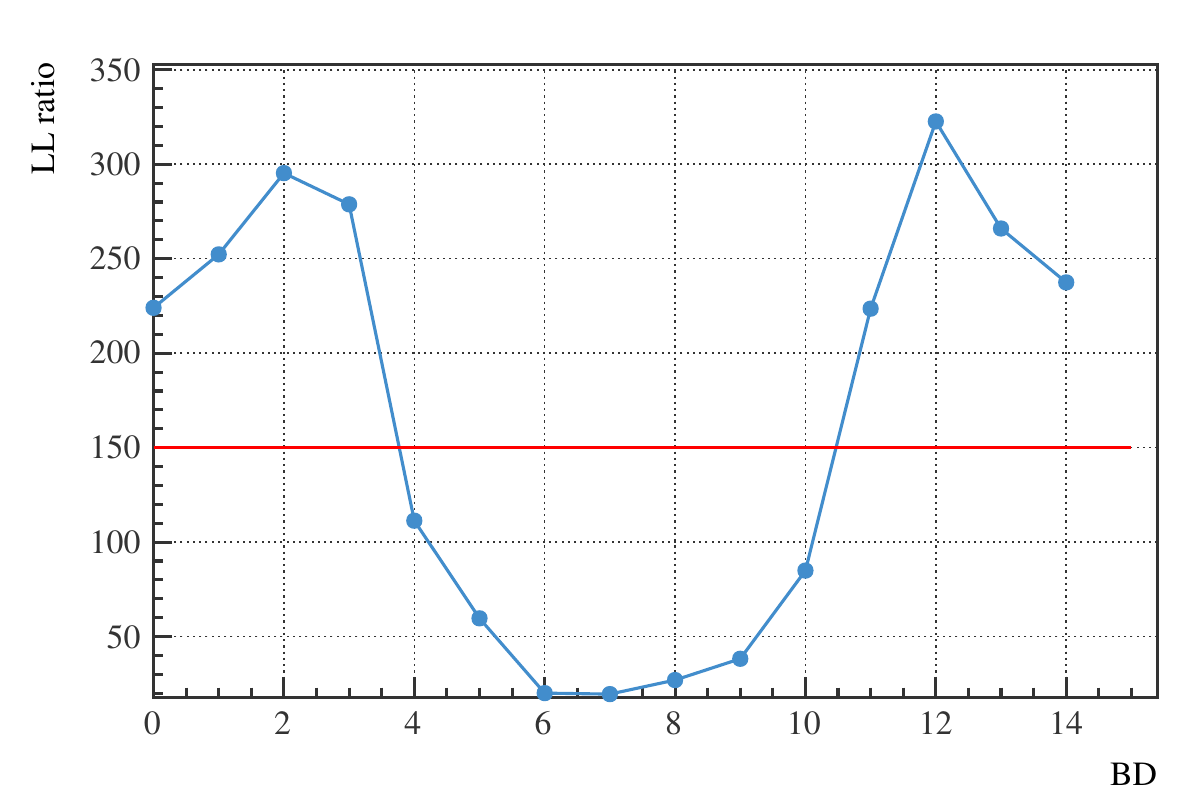}
    \caption{LL-ratios (signal+background versus background only) for crystal 3 with the $^{133}$Ba proportional response calibration. The BD number is indicated on the x-axis. BDs closer to the beamline and corresponding to lower measured recoil energies feature a smaller LL-ratio. The vertical line marks the cutoff, and any result below the line is rejected.}
    \label{ll_diff}
\end{figure}

In order to reject points where the signal cannot be resolved well, we repeat the fit for each BD and crystal with only the background model and no signal. We then calculate the Log-Likelihood ratio (LL-ratio) of the model with and without signal, and discard the QF result from the fit if the LL-ratio is smaller than a set threshold (Fig. \ref{ll_diff}). A plateau in the LL-ratio was observed for small recoil energies, as expected when the signal function is fitted to background. The Geant4 simulation already accounts for relative differences in the signal scale between datasets for different BDs. The same cutoff value for the LL-ratio is used for all crystals.

\begin{figure}[h]
    \centering
    \includegraphics[width=0.48\textwidth]{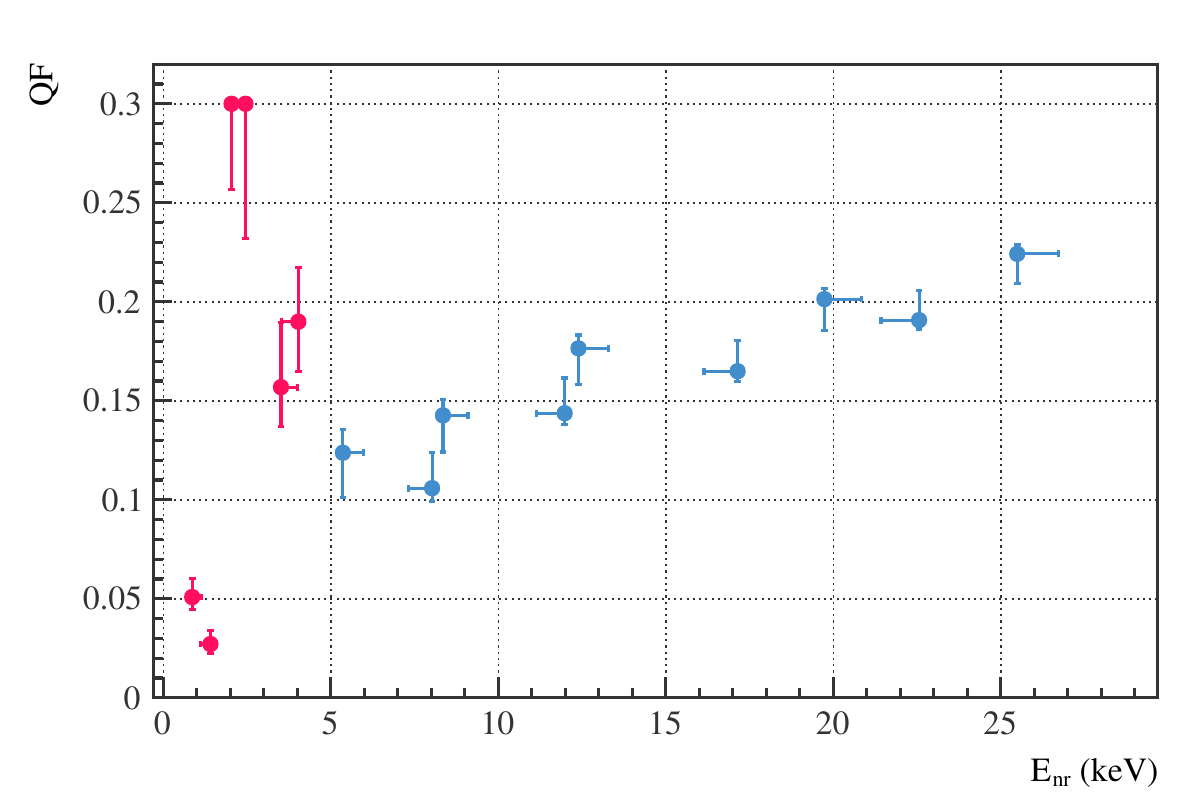}
    \caption{QF$_{\textrm{Na}}$ for Crystal 1 (using the $^{133}$Ba proportional response calibration) with rejected results from fits to background shown in red.}
    \label{LL_reject}
\end{figure}

\subsection{Geometrical uncertainites}
\label{sawtooth}
As can be seen in Fig. \ref{LL_reject}, a sawtooth pattern is present in the results. The QF seems to oscillate above and below a ``central" trend.
\begin{figure}[h]
    \centering
    \includegraphics[width=0.48\textwidth]{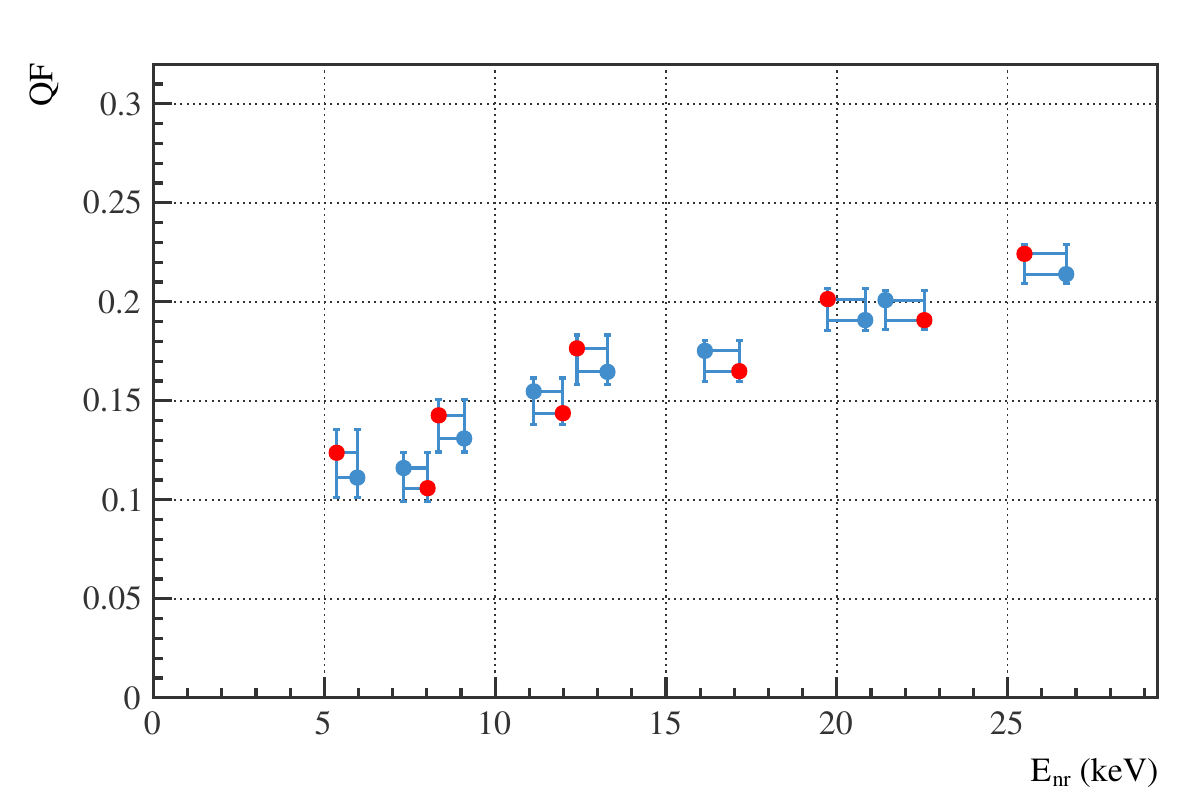}
    \caption{Red: QF$_{\textrm{Na}}$ for Crystal 1 with the sawtooth pattern; Blue: ``Flattened" QF$_{\textrm{Na}}$ and nominal recoil energies corrected for an angular shift of $1^\circ$, outlined in Sec. \ref{sawtooth}.}
    \label{sawtooth_plot}
\end{figure}
This oscillation likely stems from the alternating geometry, where the BD corresponding to the highest recoil energy is on one side, and the second highest is on the opposite side, with respect to the beam line.
We investigated whether an angular shift in the BD array, i.e., a rotation of the BD holder structure with respect to the beamline, would remove this feature.

BDs 2 and 12 on opposite sides of the array feature the highest nominal angles relative to the beamline with a difference of only $0.5^\circ$. By correcting the nominal recoil energies of these BDs such that the difference in their measured QFs becomes minimal, we obtain an angular shift of about $1^\circ$. This is calculated for each crystal, where only Crystal 4 deviates significantly, showing a smaller shift. The effect of a $1^\circ$ rotation is therefore conservatively considered as the systematic uncertainty due to a possible rotation of the holder structure. Fig. \ref{sawtooth_plot} additionally shows the QFs and nuclear recoil energies that would result for Crystal 1 from correcting for this angular rotation. As the exact angle is not known, we add an asymmetric systematic uncertainty on the assumed nuclear recoil energy for all BDs instead of applying this correction directly. This uncertainty is propagated to the QFs.


\section{Results}
\label{results}

\subsection{Energy dependence}
\begin{figure*}[!t]
    \centering
    \begin{subfigure}
        \centering
        \includegraphics[width=.49\textwidth]{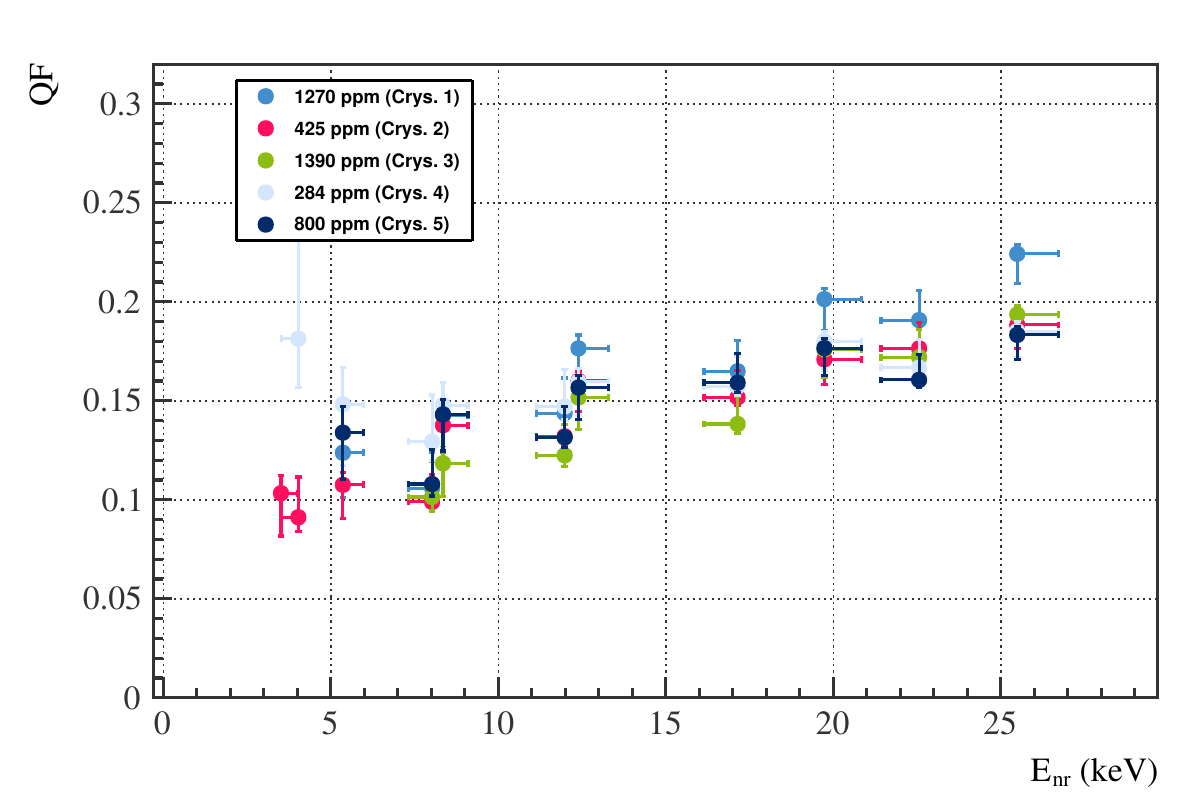}
    \end{subfigure}
    \hfill
    \begin{subfigure}
        \centering
        \includegraphics[width=.49\textwidth]{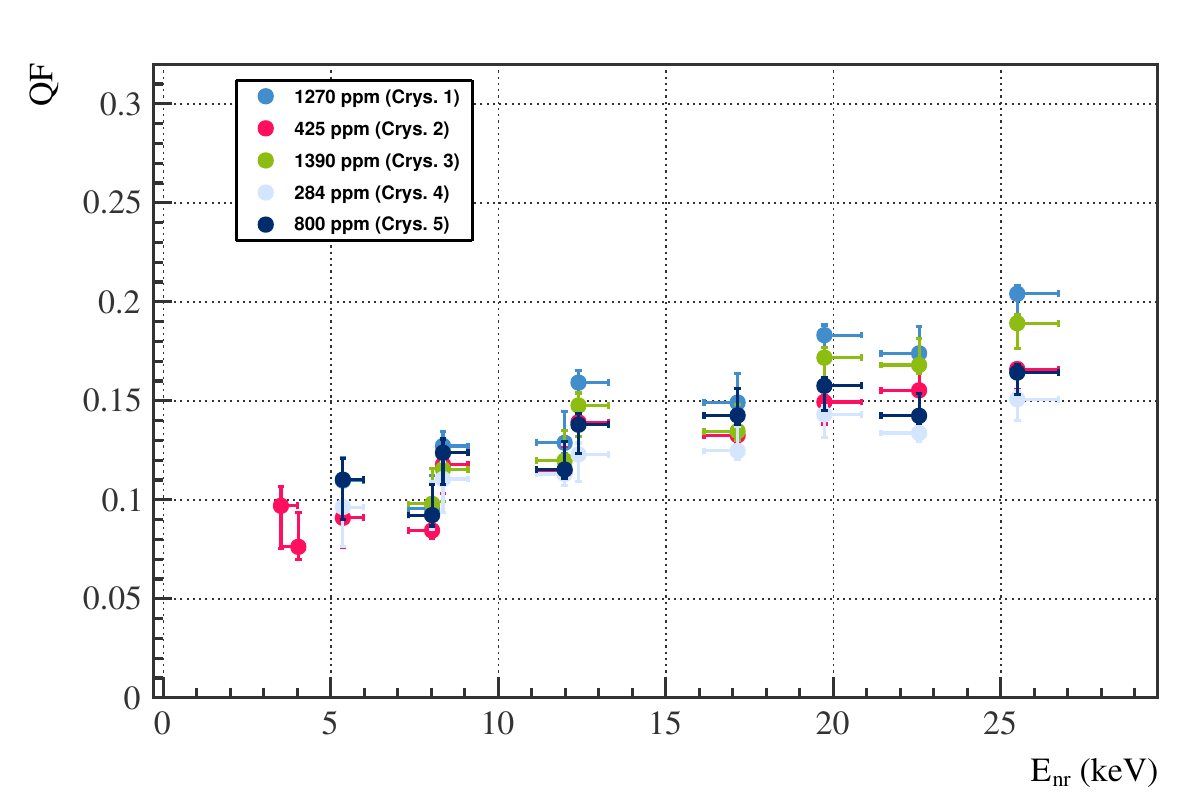}
    \end{subfigure}
    \caption{QF$_{\textrm{Na}}$ as a function of energy for calibration with $^{133}$Ba \unit[6.6]{keV} (left) and $^{241}$Am \unit[59.54]{keV} (right).}
    \label{qfs_combined}
\end{figure*}

QF$_{\textrm{Na}}$ results for the measured crystals using the methodology described in Sec. \ref{bat_description} and calibration schemes described in Sec. \ref{calibration} are shown in Fig. \ref{qfs_combined}. The uncertainties are the lower ($0.16$) and upper ($0.86$) quantiles of the posterior QF distribution, as shown in Fig. \ref{fig:posterior} for Crystal 1 and BD 0.
We find a clear energy dependence of QF$_{\textrm{Na}}$ for all five crystals, i.e., QF$_{\textrm{Na}}$ decreases with nuclear recoil energy. 

The calibration with $^{241}$Am \unit[59.54]{keV} seems to yield a slightly lower QF, consistent with the energy underestimation discussed in Sec. \ref{calibration}, and additionally causes more spread among the different crystals when compared to calibration with $^{133}$Ba \unit[6.6]{keV}.

\begin{figure}[h!]
    \centering
    \includegraphics[width=1\columnwidth]{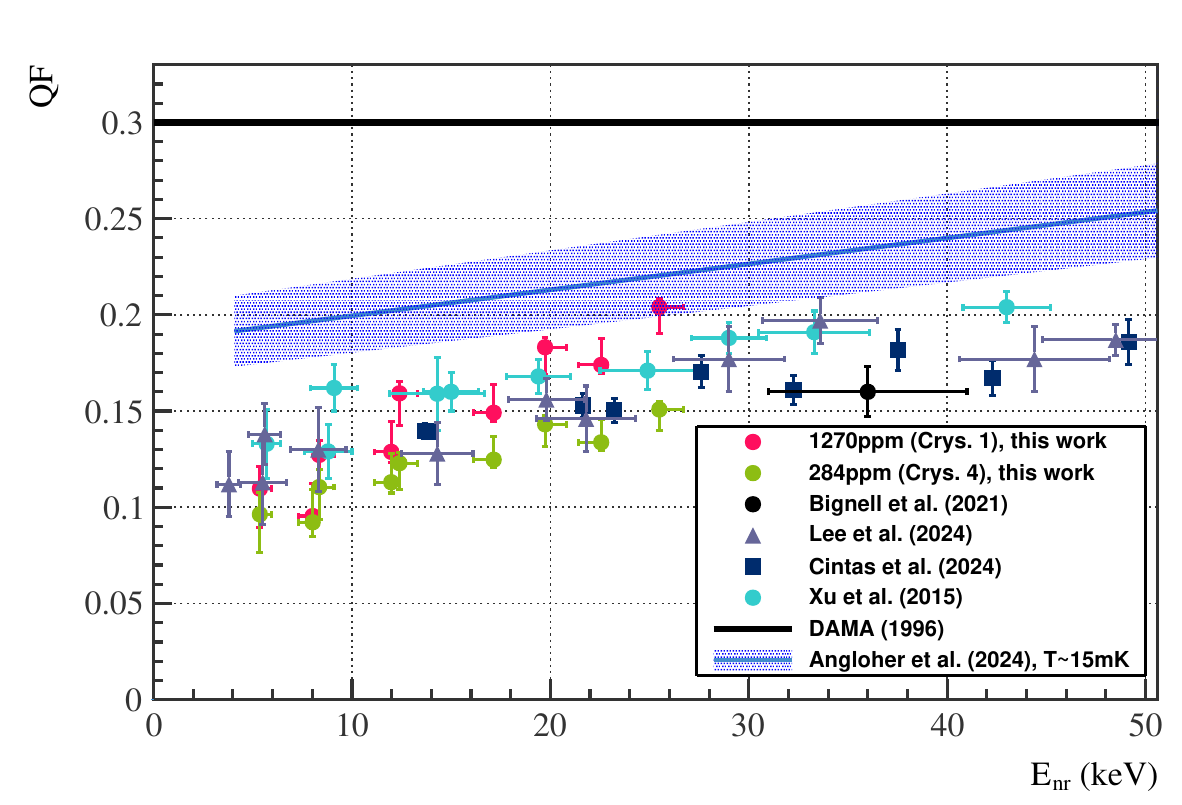}
    \caption{Comparison of QF$_{\text{Na}}$ between Crystal 1 and Crystal 4, alongside recent room temperature measurements \cite{7,8,9,10,11}, and the fit result from a cryogenic measurement \cite{summer_run}. Calibration of the room temperature data sets is based on either the \unit[57.6]{keV} $\gamma$-peak from the $^{127}$I(n,n'$\gamma$) process or the \unit[59.54]{keV}  $\gamma$-peak from $^{241}$Am.}
    \label{comparison}
\end{figure}

The lowest point we can resolve is BD5 for Crystal 2, corresponding to a recoil energy of $\unit[3.6]{keV_{\textrm{nr}}}$.
A comparison between the results of Crystal 1 and Crystal 4 with the $^{241}$Am calibration scheme and previous measurements is shown in Fig. \ref{comparison}.
All datasets yield lower QF$_{\textrm{Na}}$ at lower nuclear recoil energies.

The data from Crystal 1 data agrees reasonably well with \cite{7}, while Crystal 4 is more in line with \cite{11}. Both are compatible with \cite{10}. 
Different calibration methods or the effect of Tl dopant level could explain this behavior, which is discussed in the following section. Fig. \ref{comparison} also shows a QF fit result from a cryogenic measurement of a COSINUS prototype \cite{summer_run}, in which a Tl-doped crystal with \unit[730 $\pm$ 73]{ppm} was operated at T$\sim$\unit[15]{mK}. An empirical function QF$_{\textrm{Na}}\sim(1-a\cdot e^{-\frac{E_{\textrm{nr}}}{b}})$ was fitted to the data. The cryogenic measurement displays less quenching across the whole nuclear recoil energy range, although the Tl dopant level in the crystal is similar to the levels in this work.

\subsection{Tl dependence}
\label{tl}
\begin{figure}[h]
    \centering
    \begin{subfigure}
        \centering
        \includegraphics[width=.45\textwidth]{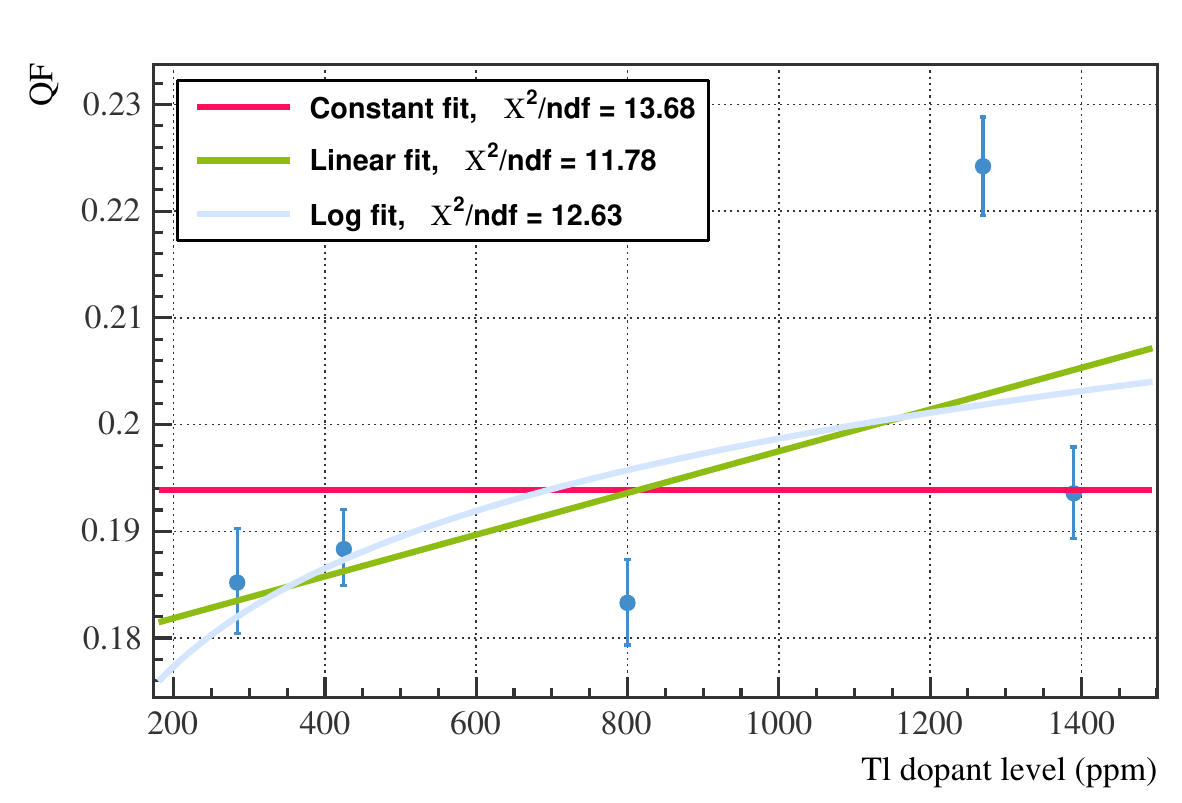}
    \end{subfigure}
    \hfill
    \begin{subfigure}
        \centering
        \includegraphics[width=.45\textwidth]{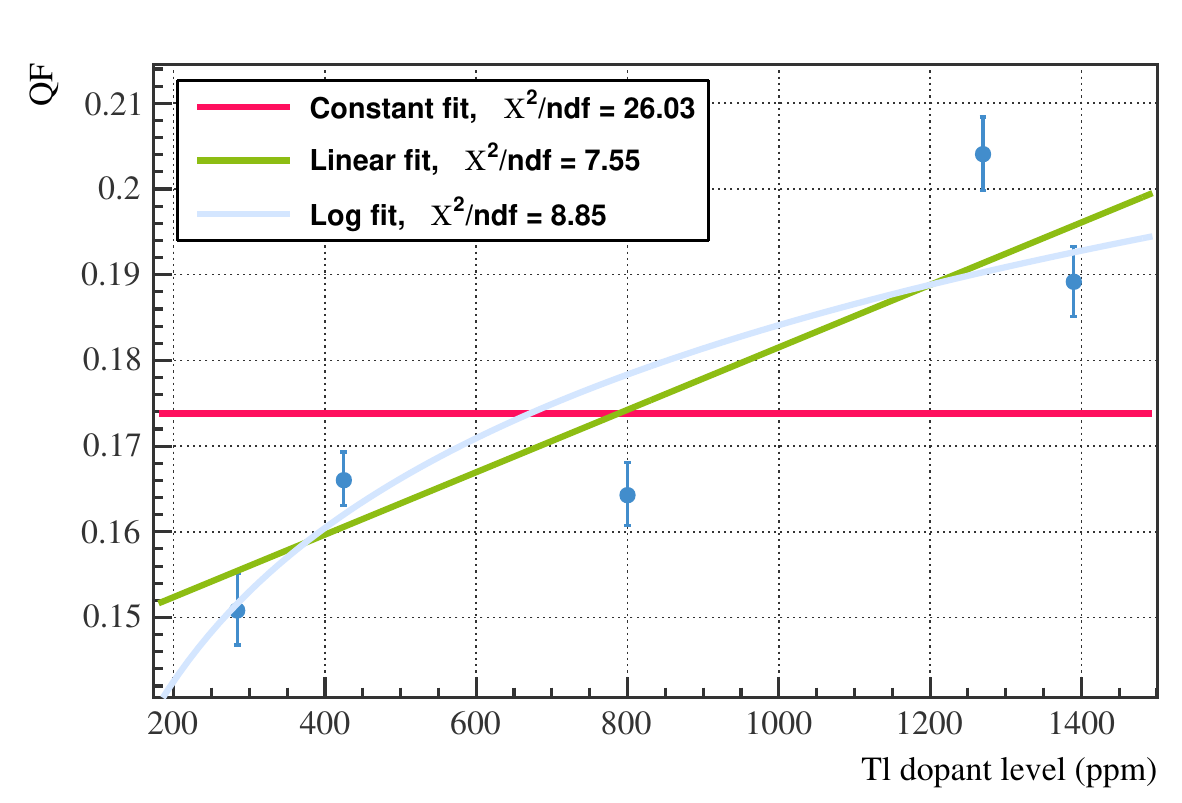}
    \end{subfigure}
    \caption{Tl dependence of QF$_{\textrm{Na}}$ for a recoil energy of \unit[$\sim$26]{keV} (corresponding to BD0) with different functional assumptions. Top: $^{133}$Ba \unit[6.6]{keV} calibration, bottom: $^{241}$Am \unit[59.54]{keV} calibration.}
    \label{Tl_dependencies_bd0}
\end{figure}

A possible Tl dependence of QF$_{\textrm{Na}}$ was investigated by selecting a specific nuclear recoil energy across all crystals, and then plotting QF$_{\textrm{Na}}$ for that energy as a function of Tl dopant level in the crystal. 
Three different fits were carried out for these plots: a constant fit, a linear fit, and a logarithmic fit of the form $a\log (x) + b$. 
By comparing the $\chi^2/n.d.f$ values of the constant fit to the linear one, we gauge whether there is a significant dependence.
The logarithmic fit results were found to be comparable to the linear case.
A comparison of the fits for a given BD energy is shown in Fig. \ref{Tl_dependencies_bd0}.

\begin{table}[b]
    \centering
    \footnotesize  
    \renewcommand{\arraystretch}{1.1}  
    \setlength{\tabcolsep}{5pt}  
    \begin{tabular}{c c c c} 
        \toprule
        \textbf{Detector no.} & \textbf{Constant} & \textbf{Linear}  & \textbf{Logarithm} \\ [0.5ex] 
        \midrule
BD0 & 13.68(26.03) & 11.78(7.55) & 12.63(8.85) \\
BD1 & 5.91(13.60) & 5.77(1.86) & 6.14(2.35) \\
BD2 & 2.20(5.24) & 2.81(2.16) & 2.82(1.84) \\
BD3 & 2.06(0.84) & 2.06(0.84) & 2.20(0.70) \\
BD11 & 2.17(1.06) & 2.86(0.43) & 2.80(0.57) \\
BD12 & 2.59(1.45) & 3.22(0.64) & 3.11(0.74) \\
BD13 & 3.92(4.33) & 4.99(2.87) & 5.09(2.13) \\
BD14 & 6.05(14.74) & 7.28(8.58) & 7.50(9.34) \\[1ex] 
        \bottomrule
    \end{tabular}
    \caption{Reduced $\chi^2$ values for the Tl-dopant fits for $^{133}$Ba \unit[6.6]{keV}($^{241}$Am \unit[59.54]{keV}) proportional response calibration.}
    \label{chisquare_tl}
\end{table}

\begin{figure}
    \centering
    \begin{subfigure}
        \centering
        \includegraphics[width=.45\textwidth]{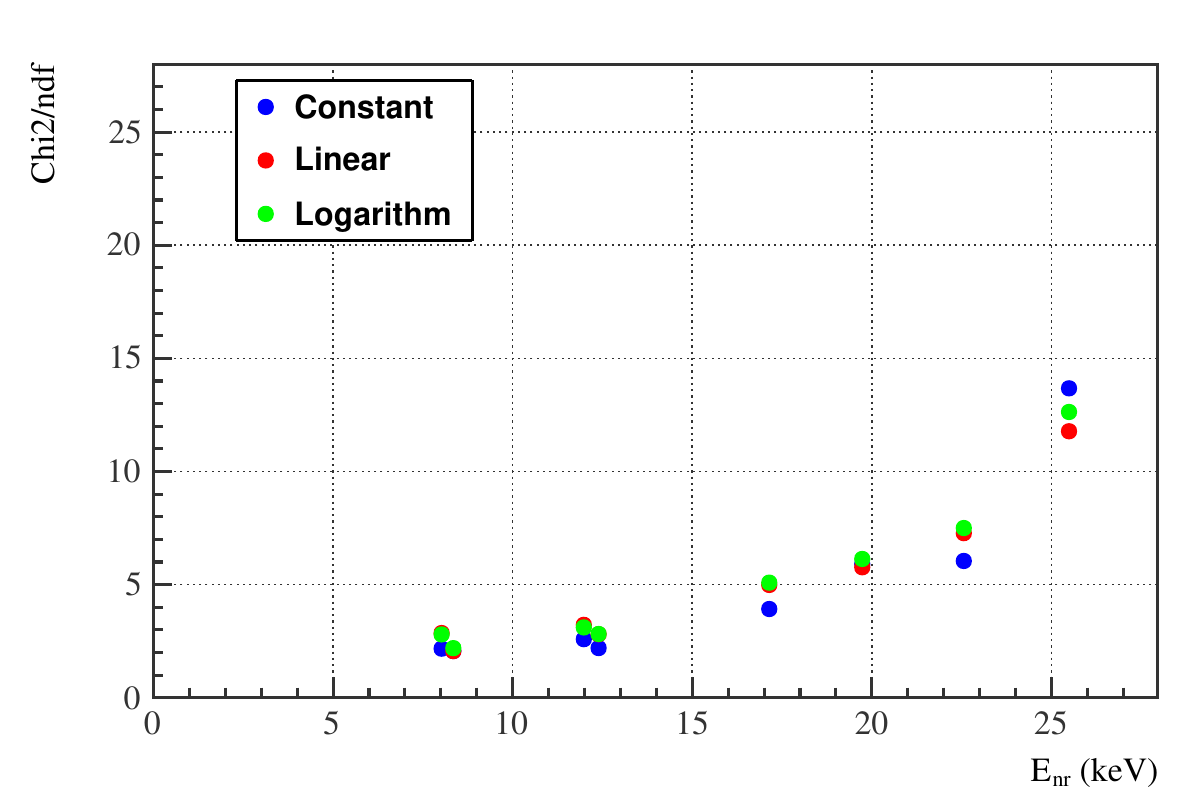}
    \end{subfigure}
    \hfill
    \begin{subfigure}
        \centering
        \includegraphics[width=.45\textwidth]{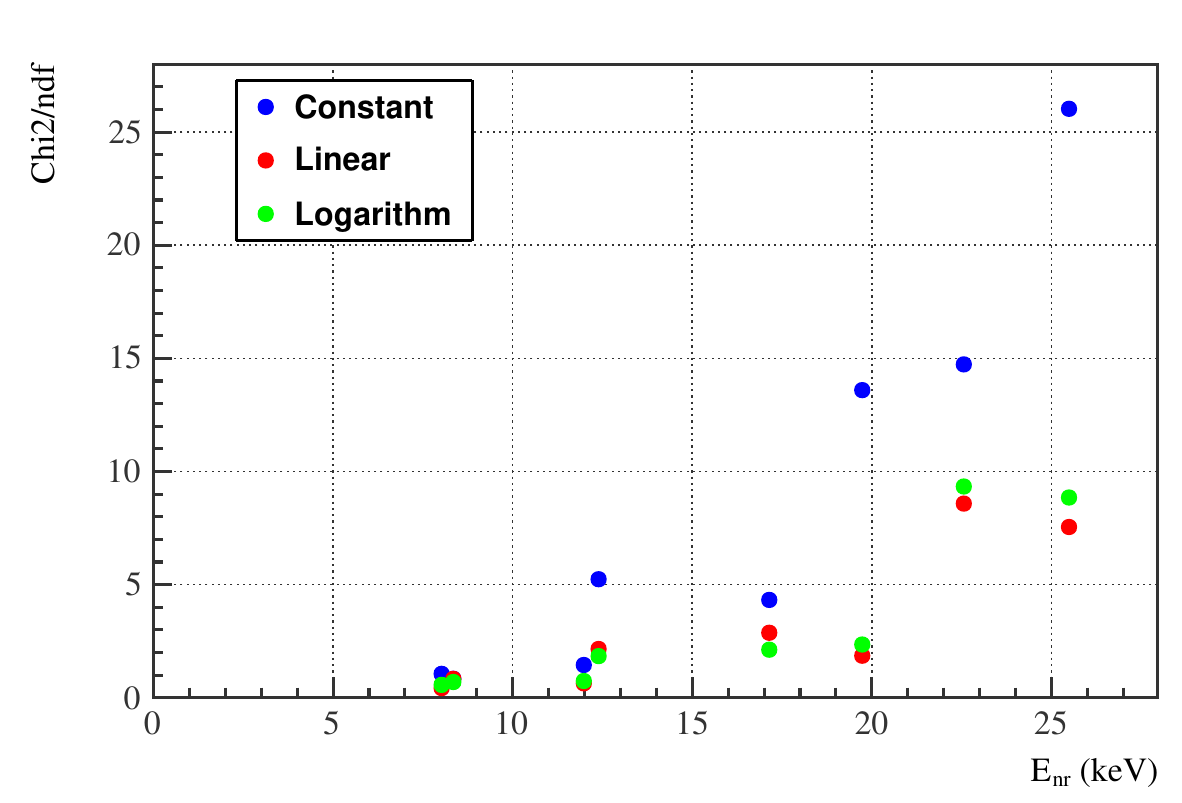}
    \end{subfigure}
    \caption{$\chi^2 /n.d.f$ values for different functional assumptions on the Tl dependence of QF$_{\textrm{Na}}$. Top: $^{133}$Ba \unit[6.6]{keV} calibration, bottom: $^{241}$Am \unit[59.54]{keV}.}
    \label{Tl_dependencies_combined}
\end{figure}

We find that the Tl dopant dependence is strongly visible for calibration with the \unit[59.54]{keV} peak from $^{241}$Am, i.e., constant QFs are a bad fit,  while it is much less pronounced for $^{133}$Ba \unit[6.6]{keV} (Fig. \ref{Tl_dependencies_combined}).
This indicates that the light yield of NaI(Tl) still has an influence on the results. However, a lower Tl dopant level seems to result in lower QF$_{\textrm{Na}}$ overall. For a majority of recoil energies, a linear dependence describes the observed data better than a logarithmic dependence.

\section{Conclusion}

Recent years have seen important advancements in the study of QF in NaI(Tl) crystals, driven by a concerted effort from the scientific community to address the long-standing uncertainty that the QF poses in the interpretation of results from NaI(Tl)-based scintillation light-only direct DM search experiments. We measured QF$_{\textrm{Na}}$ in five such crystals, each with a different Tl dopant concentration.
The recoil energies are in the range \unit[5-26]$\textrm{keV}_{\textrm{nr}}$.
Each crystal was irradiated for 28-35 hours at TUNL by a quasi-monoenergetic neutron beam.
To calibrate the crystals, $^{241}$Am and $^{133}$Ba sources were used in this work.
We calibrated the crystal response by assuming proportional response to the \unit[6.6]{keV} peak of $^{133}$Ba and the \unit[59.54]{keV} peak of $^{241}$Am.
The QF was determined through a bayesian fit using BAT, assuming a parametric background function and using quenched simulated nuclear recoil histograms for each BD, smeared with the PMT resolution function.
The QFs have a visible energy dependence and are compatible with previous results from other studies. We were able to extend the measurement to lower recoil energies. I recoils were not studied in this work due to their extremely low measured quenched energies.
A systematic geometrical effect (sawtooth pattern) is visible for all QFs, that is believed to originate from a rotation of the BD holder relative to the beam axis. This systematic uncertainty was estimated by calculating modified nuclear recoil energies with an angular shift of 1$^\circ$. The pattern has also been observed in other measurements \cite{10}.
We studied a potential Tl dependence of QF$_{\textrm{Na}}$, and find that lower Tl concentration leads to lower QF in general. However, the magnitude of this effect varies between the two calibration methods. This highlights the importance of the choice of calibration in such measurements. \\
A future study of Tl dopant levels over a wider range of concentrations while exploring a broad range of recoil energies is needed to extract a comprehensive model for the relationship between light yield and quenching characteristics.
Understanding the non-linear light yield of NaI(Tl) crystals \cite{det_cal_1,det_cal_2} is another essential aspect.

\begin{acknowledgements}
We extend our gratitude to the HEPHY/ TU Wien simulation team, especially Alexander Fuss, Samir Banik, and Holger Kluck, for their steadfast assistance in addressing several technical inquiries throughout.
This work has been supported by the Austrian Science Fund FWF, stand-alone project AnaCONDa [P 33026-N].
\end{acknowledgements}

\bibliographystyle{spphys}       
\bibliography{refs.bib}

\begin{thebibliography}{10}
\providecommand{\url}[1]{{#1}}
\providecommand{\urlprefix}{URL }
\expandafter\ifx\csname urlstyle\endcsname\relax
  \providecommand{\doi}[1]{DOI \discretionary{}{}{}#1}\else
  \providecommand{\doi}{DOI \discretionary{}{}{}\begingroup \urlstyle{rm}\Url}\fi

\bibitem{nai_discovery}
R.~Hofstadter, Phys. Rev. \textbf{75}, 796 (1949).
\newblock \doi{10.1103/PhysRev.75.796}.
\newblock \urlprefix\url{https://link.aps.org/doi/10.1103/PhysRev.75.796}

\bibitem{bernabei2022recent}
R.~Bernabei, P.~Belli, A.~Bussolotti, V.~Caracciolo, R.~Cerulli, N.~Ferrari, et~al., Moscow University Physics Bulletin \textbf{77}(2), 291 (2022).
\newblock \doi{10.15407/jnpae2021.04.329}

\bibitem{Fushimi_2016}
K.~Fushimi, et~al., Journal of Physics: Conference Series \textbf{718}, 042022 (2016).
\newblock \doi{10.1088/1742-6596/718/4/042022}.
\newblock \urlprefix\url{https://doi.org/10.1088/1742-6596/718/4/042022}

\bibitem{sabre}
M.~Antonello, et~al., Eur. Phys. J. C \textbf{79}(4), 363 (2019).
\newblock \doi{10.1140/epjc/s10052-019-6860-y}.
\newblock \urlprefix\url{https://doi.org/10.1140/epjc/s10052-019-6860-y}

\bibitem{PhysRevD.103.102005}
J.~Amar\'e, et~al., Phys. Rev. D \textbf{103}, 102005 (2021).
\newblock \doi{10.1103/PhysRevD.103.102005}.
\newblock \urlprefix\url{https://link.aps.org/doi/10.1103/PhysRevD.103.102005}

\bibitem{PhysRevD.106.052005}
G.~Adhikari, et~al., Phys. Rev. D \textbf{106}, 052005 (2022).
\newblock \doi{10.1103/PhysRevD.106.052005}.
\newblock \urlprefix\url{https://link.aps.org/doi/10.1103/PhysRevD.106.052005}

\bibitem{PhysRevD.95.032006}
E.~Barbosa~de Souza, et~al., Phys. Rev. D \textbf{95}, 032006 (2017).
\newblock \doi{10.1103/PhysRevD.95.032006}.
\newblock \urlprefix\url{https://link.aps.org/doi/10.1103/PhysRevD.95.032006}

\bibitem{billard2022direct}
J.~Billard, et~al., Reports on Progress in Physics \textbf{85}(5), 056201 (2022).
\newblock \doi{10.1088/1361-6633/ac5754}

\bibitem{Coarasa2024ANAIS112TY}
I.~Coarasa, et~al., Communications Physics \textbf{7} (2024).
\newblock \urlprefix\url{https://api.semanticscholar.org/CorpusID:269430304}

\bibitem{Carlin_2025_v2}
N.~Carlin, et~al., Science Advances \textbf{11}(36) (2025).
\newblock \doi{10.1126/sciadv.adv6503}.
\newblock \urlprefix\url{http://dx.doi.org/10.1126/sciadv.adv6503}

\bibitem{Angloher_2016}
G.~Angloher, et~al., Eur. Phys. J. C \textbf{76}, 441 (2016).
\newblock \doi{10.1140/epjc/s10052-016-4278-3}.
\newblock \urlprefix\url{https://doi.org/10.1140/epjc/s10052-016-4278-3}

\bibitem{Carlin_2025}
N.~Carlin, et~al., Physical Review Letters \textbf{135}(12) (2025).
\newblock \doi{10.1103/9j7w-qp1c}.
\newblock \urlprefix\url{http://dx.doi.org/10.1103/9j7w-qp1c}

\bibitem{1}
N.~Spooner, et~al., Phys. Lett. B \textbf{321}, 156 (1994).
\newblock \doi{10.1016/0370-2693(94)90343-3}.
\newblock \urlprefix\url{https://doi.org/10.1016/0370-2693(94)90343-3}

\bibitem{2}
R.~Bernabei, et~al., Phys. Lett. B \textbf{389}, 757 (1996).
\newblock \doi{10.1016/S0370-2693(96)80020-7}.
\newblock \urlprefix\url{https://doi.org/10.1016/S0370-2693(96)80020-7}

\bibitem{3}
G.~Gerbier, et~al., Astropart. Phys. \textbf{11}, 287 (1999).
\newblock \doi{10.1016/S0927-6505(99)00004-3}.
\newblock \urlprefix\url{https://doi.org/10.1016/S0927-6505(99)00004-3}

\bibitem{4}
E.~Simon, et~al., Nucl. Instrum. Methods A \textbf{507}, 643 (2003).
\newblock \doi{10.1016/S0168-9002(03)01438-4}.
\newblock \urlprefix\url{https://doi.org/10.1016/S0168-9002(03)01438-4}

\bibitem{5}
H.~Chagani, et~al., J. Inst. \textbf{3}, P06003 (2008).
\newblock \doi{10.1088/1748-0221/3/06/P06003}.
\newblock \urlprefix\url{https://doi.org/10.1088/1748-0221/3/06/P06003}

\bibitem{6}
J.I. Collar, Phys. Rev. C \textbf{88}, 035806 (2013).
\newblock \doi{10.1103/PhysRevC.88.035806}.
\newblock \urlprefix\url{https://doi.org/10.1103/PhysRevC.88.035806}

\bibitem{7}
J.~Xu, et~al., Phys. Rev. C \textbf{92}, 015807 (2015).
\newblock \doi{10.1103/PhysRevC.92.015807}.
\newblock \urlprefix\url{https://doi.org/10.1103/PhysRevC.92.015807}

\bibitem{8}
H.~Joo, et~al., Astropart. Phys. \textbf{108}, 50 (2019).
\newblock \doi{10.1016/j.astropartphys.2019.01.001}.
\newblock \urlprefix\url{https://doi.org/10.1016/j.astropartphys.2019.01.001}

\bibitem{9}
L.J. Bignell, et~al., Journal of Instrumentation \textbf{16}, P07034 (2021).
\newblock \doi{10.1088/1748-0221/16/07/P07034}.
\newblock \urlprefix\url{https://doi.org/10.1088/1748-0221/16/07/P07034}

\bibitem{10}
D.~Cintas, et~al., Phys. Rev. C \textbf{110}, 014613 (2024).
\newblock \doi{10.1103/PhysRevC.110.014613}.
\newblock \urlprefix\url{https://link.aps.org/doi/10.1103/PhysRevC.110.014613}

\bibitem{11}
S.H. Lee, et~al., Phys. Rev. C \textbf{110}, 014614 (2024).
\newblock \doi{10.1103/PhysRevC.110.014614}.
\newblock \urlprefix\url{https://link.aps.org/doi/10.1103/PhysRevC.110.014614}

\bibitem{christmas_run}
G.~Angloher, et~al., Phys. Rev. D \textbf{109}, 082003 (2024).
\newblock \doi{10.1103/PhysRevD.109.082003}.
\newblock \urlprefix\url{https://link.aps.org/doi/10.1103/PhysRevD.109.082003}

\bibitem{summer_run}
G.~Angloher, et~al., Phys. Rev. D \textbf{110}, 043010 (2024).
\newblock \doi{10.1103/PhysRevD.110.043010}.
\newblock \urlprefix\url{https://link.aps.org/doi/10.1103/PhysRevD.110.043010}

\bibitem{denis}
G.~Lawrence, R.~Beauchamp, J.~McKibben, Nucl. Instrum. and Meth. \textbf{32}, 357 (1965).
\newblock \doi{10.1016/0029-554X(65)90539-2}.
\newblock \urlprefix\url{https://doi.org/10.1016/0029-554X(65)90539-2}

\bibitem{LISKIEN197557}
H.~Liskien, A.~Paulsen, Atomic Data and Nuclear Data Tables \textbf{15}(1), 57 (1975).
\newblock \doi{https://doi.org/10.1016/0092-640X(75)90004-2}.
\newblock \urlprefix\url{https://www.sciencedirect.com/science/article/pii/0092640X75900042}

\bibitem{FRIEDMAN2013117}
M.~Friedman, et~al., Nucl. Instrum. and Meth. Phys. A \textbf{698}, 117 (2013).
\newblock \doi{https://doi.org/10.1016/j.nima.2012.09.027}.
\newblock \urlprefix\url{https://www.sciencedirect.com/science/article/pii/S0168900212010820}

\bibitem{siccas}
Y.~Zhu, et~al., \emph{{Production of ultra-low radioactivity NaI (Tl) crystals for Dark Matter detectors}} (2018), pp. 1--3.
\newblock \doi{10.1109/NSSMIC.2018.8824322}

\bibitem{merck}
https://www.merckgroup.com/  (2024).
\newblock \urlprefix\url{https://www.merckgroup.com/en}

\bibitem{LNGS}
S.~Nisi, L.~Copia, I.~Dafinei, M.L. Di~Vacri, Int. J. of Modern Phys. A \textbf{32}, 1743003 (2017).
\newblock \doi{10.1142/S0217751X17430035}

\bibitem{eljen}
A.~Tomanin, et~al., Nucl. Instrum. and Meth. Phys. A \textbf{756}, 45 (2014).
\newblock \doi{https://doi.org/10.1016/j.nima.2014.03.028.}
\newblock \urlprefix\url{https://www.sciencedirect.com/science/article/pii/S0168900214003222}

\bibitem{ngmdaq}
{A toolkit for data acquisition control and analysis for waveform digitizers including GAGE, XIA, and Struck}.
\newblock \url{https://code.ornl.gov/CASA/ngmdaq}

\bibitem{det_cal_1}
D.W. Aitken, B.L. Beron, G.~Yenicay, H.R. Zulliger, IEEE Transactions on Nuclear Science \textbf{14}(1), 468 (1967).
\newblock \doi{10.1109/TNS.1967.4324457}

\bibitem{det_cal_2}
B.~Rooney, J.~Valentine, IEEE Transactions on Nuclear Science \textbf{44}(3), 509 (1997).
\newblock \doi{10.1109/23.603702}

\bibitem{refId0}
{COSINE-100 Collaboration}, et~al., Eur. Phys. J. C \textbf{84}(5), 484 (2024).
\newblock \doi{10.1140/epjc/s10052-024-12770-1}.
\newblock \urlprefix\url{https://doi.org/10.1140/epjc/s10052-024-12770-1}

\bibitem{GEANT4:2002zbu}
S.~Agostinelli, et~al., Nucl. Instrum. Meth. A \textbf{506}, 250 (2003).
\newblock \doi{10.1016/S0168-9002(03)01368-8"}

\bibitem{Allison2006Geant}
J.~Allison, et~al., IEEE Transactions on Nuclear Science \textbf{53}(1), 270 (2006).
\newblock \doi{10.1109/TNS.2006.869826}.
\newblock \urlprefix\url{https://doi.org/10.1109/TNS.2006.869826}

\bibitem{ALLISON2016186}
J.~Allison, et~al., {Nucl. Instrum. and Meth. A} \textbf{835}, 186 (2016).
\newblock \doi{https://doi.org/10.1016/j.nima.2016.06.125}.
\newblock \urlprefix\url{https://www.sciencedirect.com/science/article/pii/S0168900216306957}

\bibitem{CRESST:2019oqe}
A.H. Abdelhameed, et~al., Eur. Phys. J. C \textbf{79}(10), 881 (2019).
\newblock \doi{10.1140/epjc/s10052-019-7504-y}.
\newblock \urlprefix\url{https://doi.org/10.1140/epjc/s10052-019-7504-y}

\bibitem{abdelhameed2019first}
A.H. Abdelhameed, et~al., Physical Review D \textbf{100}(10), 102002 (2019).
\newblock \doi{10.1103/PhysRevD.100.102002}.
\newblock \urlprefix\url{https://doi.org/10.1103/PhysRevD.100.102002}

\bibitem{brun1997root}
R.~Brun, F.~Rademakers, Nucl. Instrum. and Meth. A \textbf{389}(1-2), 81 (1997).
\newblock \doi{10.1016/S0168-9002(97)00048-X}.
\newblock \urlprefix\url{https://doi.org/10.1016/S0168-9002(97)00048-X}

\bibitem{srim}
J.~Ziegler.
\newblock {SRIM code}.
\newblock \url{http://www.srim.org}.
\newblock \textit{Last accessed}: Nov 19th, 2025

\bibitem{Caldwell_2009}
A.~Caldwell, D.~Kollár, K.~Kröninger, Computer Physics Communications \textbf{180}(11), 2197–2209 (2009).
\newblock \doi{10.1016/j.cpc.2009.06.026}.
\newblock \urlprefix\url{http://dx.doi.org/10.1016/j.cpc.2009.06.026}

\bibitem{brooks_gelman}
S.~Brooks, A.~Gelman, J. Comput. Graphi. Stat. \textbf{7}, 434 (1998).
\newblock \doi{10.1080/10618600.1998.10474787}

\end{thebibliography}

\end{document}